\documentclass[lettersize,journal]{IEEEtran}
\usepackage{amsmath,amsfonts}
\usepackage{algorithmic}
\usepackage{array}
\usepackage{textcomp}
\usepackage{stfloats}
\usepackage{url}
\usepackage{verbatim}
\usepackage{graphicx}
\usepackage{multirow}
\usepackage{booktabs}
\usepackage{makecell}
\usepackage{xcolor}
\usepackage{hyperref}
\usepackage{subcaption}
\usepackage[numbers, sort&compress]{natbib}

\makeatletter
\newcommand\footnoteref[1]{\protected@xdef\@thefnmark{\ref{#1}}\@footnotemark}
\makeatother

\def\BibTeX{{\rm B\kern-.05em{\sc i\kern-.025em b}\kern-.08em
    T\kern-.1667em\lower.7ex\hbox{E}\kern-.125emX}}
\usepackage{balance}
\begin{document}
\title{An Investigation of Time-Frequency Representation Discriminators for High-Fidelity Vocoder}
\author{Yicheng Gu, \textit{Student Member, IEEE}, Xueyao Zhang, Liumeng Xue, \textit{Member, IEEE} \\ Haizhou Li, \textit{Fellow, IEEE}, Zhizheng Wu, \textit{Senior Member, IEEE}}

\maketitle

\begin{abstract}

Generative Adversarial Network (GAN) based vocoders are superior in both inference speed and synthesis quality when reconstructing an audible waveform from an acoustic representation. This study focuses on improving the discriminator for GAN-based vocoders. Most existing Time-Frequency Representation (TFR)-based discriminators are rooted in Short-Time Fourier Transform (STFT), which owns a constant Time-Frequency (TF) resolution, linearly scaled center frequencies, and a fixed decomposition basis, making it incompatible with signals like singing voices that require dynamic attention for different frequency bands and different time intervals. Motivated by that, we propose a Multi-Scale Sub-Band Constant-Q Transform CQT (MS-SB-CQT) discriminator and a Multi-Scale Temporal-Compressed Continuous Wavelet Transform CWT (MS-TC-CWT) discriminator. Both CQT and CWT have a dynamic TF resolution for different frequency bands. In contrast, CQT has a better modeling ability in pitch information, and CWT has a better modeling ability in short-time transients. Experiments conducted on both speech and singing voices confirm the effectiveness of our proposed discriminators. Moreover, the STFT, CQT, and CWT-based discriminators can be used jointly for better performance. The proposed discriminators can boost the synthesis quality of various state-of-the-art GAN-based vocoders, including HiFi-GAN, BigVGAN, and APNet.


\end{abstract}


\section{Introduction}

A vocoder has always been important for various audio generation tasks (e.g., Singing Voice Synthesis (SVS)~\cite{POPCS, hiddensinger}, Text-To-Speech (TTS)~\cite{fs2, ns2}). It reconstructs an audible waveform from an acoustic feature~\cite{fs2, encodec, funcodec} outputted by the acoustic model, directly affecting the resulting audio quality and generation speed. Among different types of vocoders, the neural network-based ones are essential due to their superior synthesis quality compared to the DSP-based ones~\cite{straight, world}.

The synthesis quality and the inference speed are two primary considerations when designing a neural vocoder. Initially, autoregressive-based vocoders, represented by WaveNet~\cite{WaveNet} and WaveRNN~\cite{WaveRNN}, first successfully model the speech signal via deep neural networks. Although the autoregressive-based vocoders achieved breakthroughs regarding synthesis quality, their inference speed cannot meet the need for real-world applications due to the degradation brought by the sample-by-sample generation scheme. Subsequently, distilling-based models (e.g., ParallelWaveNet~\cite{parallelwavenet}, ClariNet~\cite{clarinet}), flow-based models (e.g., WaveFlow~\cite{WaveFlow}, WaveGlow~\cite{WaveGlow}), glottis-based models (e.g., GlotNet~\cite{glotnet}, LPCNet~\cite{lpcnet}t), diffusion-based models (e.g., DiffWave~\cite{DiffWave}, FreGrad~\cite{fregrad}), and differentialble digital signal processing (DDSP)-based models (e.g., NSF~\cite{NSF}, GOLF~\cite{golf}) were proposed. Although the inference efficiency for these models is significantly boosted, the synthesis quality was relatively degraded, making them incompatible with applications that require a high synthesis quality. Regarding this matter, GAN-based vocoders~\cite{PWG, MelGAN, UniversalMelGAN, HiFiGAN, FreGAN, SingGAN, BigVGAN, bigvsan} are proposed and currently widely used. Specifically, the parallelized generation scheme via noncausal convolution layers ensures the inference speed, and the adversarial training, which makes it possible to utilize audio-level losses in different perspectives, ensures the synthesis quality. However, to synthesize expressive speech or singing voice, current GAN-based vocoders still hold problems like spectral artifacts such as \textbf{\textit{hissing noise}}~\cite{FreGAN}, \textbf{\textit{glitches}}~\cite{sawsing}, and \textbf{\textit{the loss of details in mid and low-frequency regions}}~\cite{SingGAN}.

To pursue high-quality GAN-based vocoders, the existing studies aim to improve both the generator and the discriminator. For the generator, SingGAN~\cite{SingGAN} adopts a neural source filter (NSF)~\cite{NSF} module to utilize the phase-continuous sine excitation to alleviate the glitches~\cite{sawsing}. BigVGAN~\cite{BigVGAN} introduces a new activation function~\cite{snake} with anti-aliasing modules to improve the generalization ability and the reconstruction quality in high-frequency bands. SnakeGAN~\cite{snakegan} utilizes a fast DDSP vocoder~\cite{sawsing} to generate raw audio for providing prior knowledge. For the discriminator, MelGAN~\cite{MelGAN} employs a time-domain-based discriminator that successfully models waveform structures at different scales for the first time. HiFi-GAN~\cite{HiFiGAN} extends it with a Multi-Scale Discriminator (MSD) and Multi-Period Discriminator (MPD), which operates on the reshaped audio signal obtained from average pooling and periodical sampling individually. Fre-GAN~\cite{FreGAN} further improves them by replacing the pooling and the sampling process with Discrete Wavelet Transform (DWT) to avoid aliasing effects. UniversalMelGAN~\cite{UniversalMelGAN} introduces a Multi-Resolution Discriminator that operates on the amplitude spectrogram of an STFT matrix, followed by~\cite{MRD} emphasizing its significance in boosting the synthesis quality. Furthermore, Encodec~\cite{encodec} extends it to the Multi-Scale STFT (MS-STFT) Discriminator by utilizing the phase information together.


Among the existing works, most Time-Frequency Representation (TFR)-based discriminators~\cite{UniversalMelGAN, encodec, MRD} are rooted in the Short-Time Fourier Transform (STFT)~\cite{STFT}, which could fast extract easy-to-handle STFT spectrograms for neural networks. However, an STFT spectrogram has a constant Time-Frequency (TF) resolution, linearly scaled center frequencies, and a fixed decomposition basis. \textit{When encountering signals like singing voices, which require dynamic attention for different frequency bands and different time intervals~\cite{SingGAN}, only an STFT spectrogram will be insufficient}.

This study focuses on improving the discriminator part to improve the synthesis quality. In particular, we utilize the Constant-Q Transform (CQT)~\cite{CQT1992, CQT2010} and the Continuous Wavelet Transform (CWT)~\cite{CWT1992} to design discriminators. Both CQT and CWT have a flexible resolution for different frequency bands, which brings a better modeling ability in F0 accuracy and harmonic tracking. Moreover, CQT has a better modeling ability in pitch-level information, and CWT has a better modeling ability in short-time transients. To make CQT and CWT spectrograms feasible with the GAN-based framework, we propose the Sub-Band Processor module and the Temporal Compressor module. Based on them, we design a Multi-Scale Sub-Band CQT (MS-SB-CQT) Discriminator~\cite{CQTICASSP} and a Multi-Scale Temporal-Compressed CWT (MS-TC-CWT) Discriminator that operate on CQT and CWT spectrograms in different scales individually. The main contributions of this paper are summarized as follows: 
\begin{itemize}
    \item To make the CQT feasible with the discriminator, we propose a Sub-Band Processor for CQT to tackle the temporal desynchronization issue in the CQT spectrogram.
    \item To make the CWT feasible with the GAN-based framework, we propose a Temporal Compressor for CWT to compress the high-dimensional CWT spectrogram into the low-dimensional latent representation.
    \item To increase the diversity of the discriminated features, we modify the Multi-Resolution Processing for CQT and CWT and propose a Multi-Basis Processing technique to integrate CWT into the same Multi-Scale Processing framework based on multiple sub-discriminators.
    \item To utilize the complementary role between different TFRs, we present a joint training strategy to use multiple discriminators based on STFT, CQT, and CWT.
\end{itemize}

\textbf{\textit{To the best of our knowledge, this is the first study that employs multiple TF analysis techniques in a single GAN-based vocoder framework.}} Our proposed framework can improve the synthesis quality in the spectrogram and pitch stability without impacting the inference stage of the generator.




\section{Background: Time-Frequency Analysis}
\label{sec:related-work}


\begin{table}[tp]
\centering
\caption{A comparison of the Short-Time Fourier Transform (STFT), Constant-Q Transform (CQT), and Continuous Wavelet Transform (CWT) in terms of decomposition basis, frequency distribution, and TF resolution.}\label{tab:three-transforms}
\resizebox{\linewidth}{!}{%
\begin{tabular}{cllllll}
\toprule
\textbf{Transform} & \textbf{TF Resolution} & \textbf{Basis} & \textbf{Frequency Distribution} 
\\ \midrule
\textbf{STFT}~\cite{STFT} & \makecell[l]{Constant} & \makecell[l]{Fourier} & \makecell[l]{Arithmetic Series} 
\\ \midrule
\textbf{CQT}~\cite{CQT2010} & \makecell[l]{Dynamic} & \makecell[l]{Fourier} & \makecell[l]{Geometric Series} 
\\ \midrule
\textbf{CWT}~\cite{CWT1992} & \makecell[l]{Dynamic} & \makecell[l]{Wavelet} & \makecell[l]{Harmonic Series} 
\\ \bottomrule
\end{tabular}%
}
\end{table}

Time-Frequency (TF) analysis aims to convert a time-domain signal into a TFR over both time and frequency domains. In this section, we will briefly introduce three classical TF analysis techniques: the Short-Time Fourier Transform (STFT), Constant-Q Transform (CQT), and Continuous Wavelet Transform (CWT). A comparison of the three transforms is presented in Table~\ref{tab:three-transforms}, 





\begin{figure}[tp]
    \centering
    \begin{subfigure}[b]{0.49\columnwidth}
         \centering
         \includegraphics[width=\columnwidth]{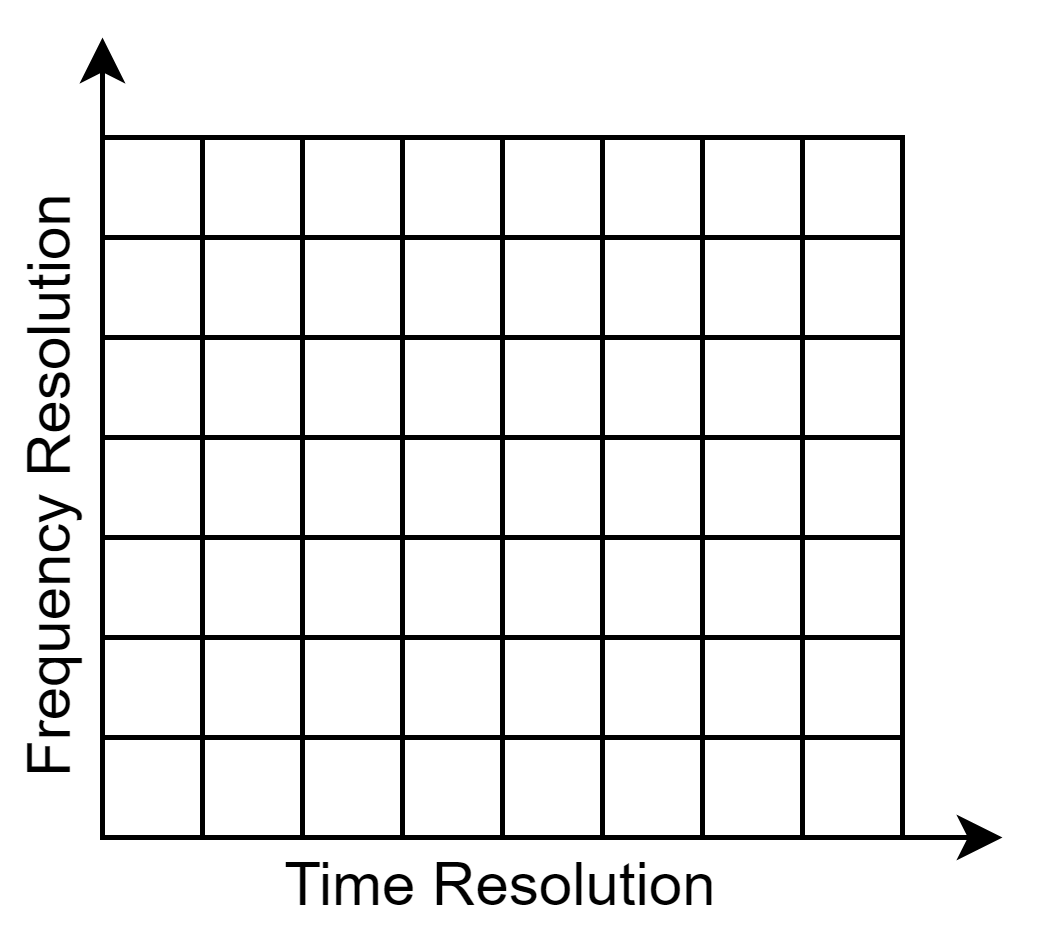}
         \caption{STFT}
         \label{fig:tf-stft}
    \end{subfigure}
    \hfill
    \begin{subfigure}[b]{0.49\columnwidth}
         \centering
         \includegraphics[width=\columnwidth]{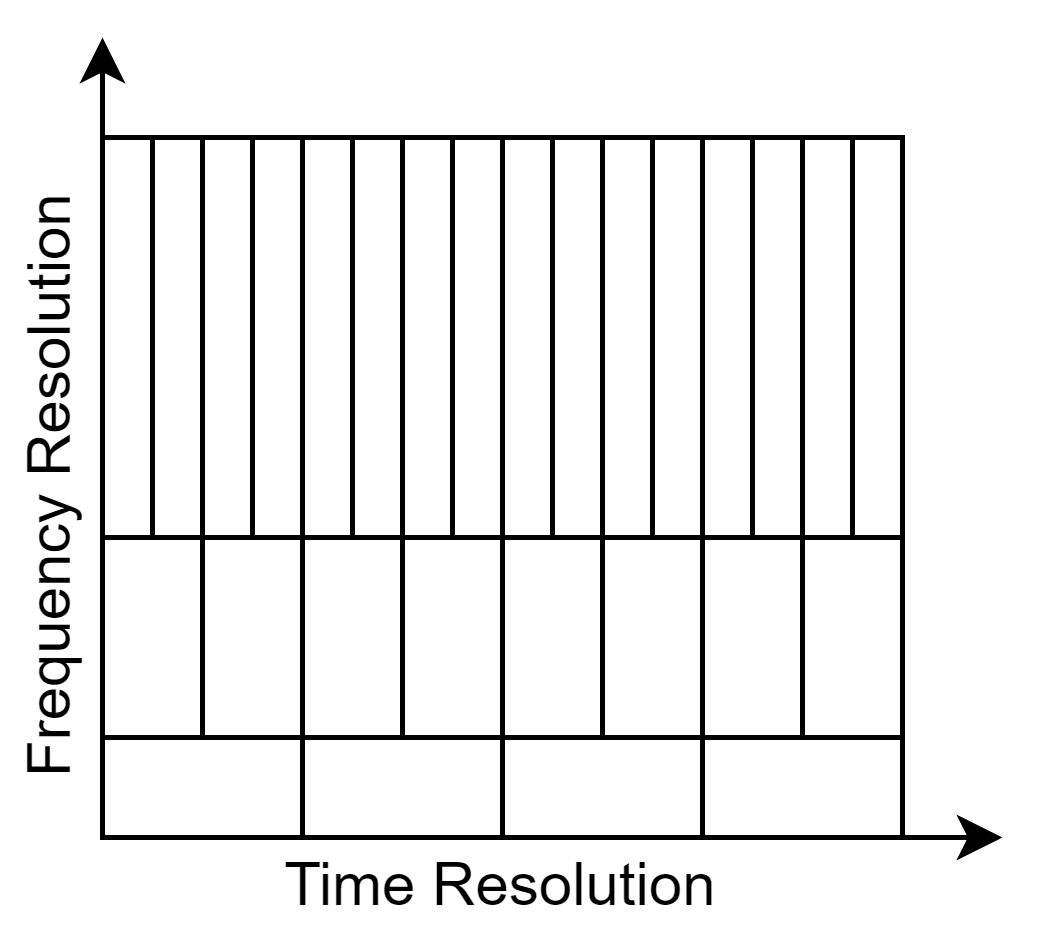}
         \caption{CQT and CWT}
         \label{fig:tf-cqt}
    \end{subfigure}
    \caption{Illustration of the TF resolution of the STFT, CQT, and CWT. A \textit{thinner} chunk in the time/frequency axis means a better time/frequency resolution. It can be observed that the CQT and CWT spectrogram have a higher frequency resolution in low-frequency bands and a higher time resolution in high-frequency bands, while the STFT spectrogram has a fixed TF resolution across all frequency bands.}
    \label{fig:resolution}
\end{figure}

\subsection{Short-Time Fourier Transform (STFT)}


Short-Time Fourier Transform (STFT)~\cite{STFT} converts a time-domain signal $x$ into its TFR, as:
\begin{equation}
\begin{split}
\label{eq:stft}
    \boldsymbol{X} ^ {\text{STFT}} (k, n) = \sum_{j = 0} ^ {N _ {k}} x (j + n - \lfloor N _ {k} / 2 \rfloor) w \left(j\right) e ^ {-i2\pi j \frac{Q _ {k}}{N _ {k}}},
\end{split}
\end{equation}
where $k$ is the index of frequency bin, $N _ {k}$ is the window length, $x(t)$ denotes the $t$-th sample point, $w(t)$ is the window function, and $Q _ {k}$ is the Q-factor, which is defined as,
\begin{equation}
\begin{split}
\label{eq:Qfactor-def}
    Q _ {k} \overset{\textit{ref.}}{=} \frac{f _ {k}}{\Delta f _ {k}},
\end{split}
\end{equation}
where $f _ {k}$ is the center frequency, $\Delta f _ {k}$ is the bandwidth. The center frequency $f _ {k}$ can be obtained as:

\begin{equation}
\begin{split}
\label{eq:stft-center-freq}
    f _ {k} = \frac{k f _ {s}}{N _ {fft}},
\end{split}
\end{equation}
where $N _ {fft}$ is the number of FFT bins. The bandwidth $\Delta f _ {k}$, which determines the TF resolution trade-off, is defined as:

\begin{equation}
\begin{split}
\label{eq:bandwidth-def}
    \Delta f _ {k} = \frac{f _ {s}}{N _ {k}},
\end{split}
\end{equation}
where $f _ {s}$ is the sampling rate.

Although STFT can be easily implemented to model a speech signal, it also has drawbacks when encountering expressive speech or singing voice due to its characteristics (Table~\ref{tab:three-transforms}), which are listed as follows:

\begin{itemize}
    \item \textbf{Fixed TF resolution:} In STFT, the window length $N _ {k}$ is fixed, bringing the $\Delta f _ {k}$ a constant (Eq.~\ref{eq:bandwidth-def}). This means the TF resolution is fixed for all frequency bins. Thus, STFT is not a good candidate to model signals that require a dynamic resolution for different frequency bands.
    \item \textbf{Limitation in harmonics modeling:} The center frequency $f _ {k}$s are linearly distributed in STFT (Eq.~\ref{eq:stft-center-freq}), which is incompatible with signals made up of harmonic frequency components that require geometrically distributed center frequencies for accurate modeling.
    \item \textbf{Inaccurate modeling of short-time transients:} According to the Gibbs phenomenon~\cite{gibbs}, when converging a square wave using the decomposition basis $e ^ {-i 2\pi n k}$ in STFT, no matter how many harmonics are used, a non-neglectable approximation error will always exist, meaning a poor modeling ability in short-time transients.
\end{itemize}




\subsection{Constant-Q Transform (CQT)}

Constant-Q Transform (CQT)~\cite{CQT1992} converts a time-domain signal $x$ into its TFR with a constant Q-factor, as:
\begin{equation}
\begin{split}
\label{eq:cqt}
    \boldsymbol{X} ^ {CQT} (k, n) = \sum_{j = n - \lfloor N _ {k} / 2 \rfloor }^{n + \lfloor N _ {k} / 2 \rfloor} x (j) a _ {k} ^ \ast (j - n + N _ {k} / 2),
\end{split}    
\end{equation}
where $a_{k} (n)$ is a complex-valued convolutional kernel, and $a _ {k} ^ \ast (n)$ is the complex conjugate of $a _ {k} (n)$. $\lfloor \cdot \rfloor$ denotes rounding down. The kernels $a _ {k} (n)$ can be obtained as:
\begin{equation}
\label{eq:cqt-kernel}
    a _ {k} (n) = \frac{1}{N _ {k}} w \left(\frac{n}{N _ {k}}\right) e ^ {-i2 \pi n \frac{Q _ {k}}{N _ {k}}},
\end{equation}
where the quality factor $Q _ {k}$s are defined constantly to conform with the human auditory system~\cite{constantQ1, constantQ2}:
\begin{equation}
\label{eq:cqt-Qfactor}
    Q _ {k} \overset{\textit{ref.}}{=} \frac{f _ {k}}{\Delta f _ {k}} = (2 ^ {\frac{1}{B}} - 1) ^ {-1},
\end{equation}
where $B$ is the number of bins per octave. The center frequencies $f_{k}$ in CQT are defined as:
\begin{equation}
\label{eq:center_freq}
\begin{split}
    f _ {k} = f _ {1} \cdot 2 ^ {\frac{k - 1}{B}},
\end{split}
\end{equation}
where $f _ {1}$ is the lowest center frequency.

Compared with STFT, CQT overcomes the fixed TF resolution and the linearly-distributed center frequencies (Table~\ref{tab:three-transforms}), resulting in the following advantages:

\begin{itemize}
    \item \textbf{Dynamic TF resolution:} CQT utilizes a constant Q-factor (Eq.~\ref{eq:cqt-Qfactor}) to obtain a dynamic TF resolution across different frequency bins. Thus, as illustrated in Fig.~\ref{fig:resolution}, the low-frequency bands will have a smaller $\Delta f _ {k}$, bringing a higher \textit{frequency resolution}, which could model the F0 more accurately; The high-frequency bands will have a bigger $\Delta f _ {k}$, bringing a higher \textit{time resolution}, which could track fast-changing harmonics variations better.
    \item \textbf{Enhanced capability in harmonic modeling:} For a better modeling ability with the sound that is made up of harmonic components, CQT utilizes a series of geometrically distributed center frequencies (Eq.~\ref{eq:center_freq}). In our study, we set $f _ {1}$ to 32.7 Hz (C1) to ensure the center frequencies conform with the notes in Western Music. 
\end{itemize}

\begin{figure}[htp!]
    \centering
    \begin{subfigure}[b]{0.45\columnwidth}
         \centering
         \includegraphics[width=\columnwidth]{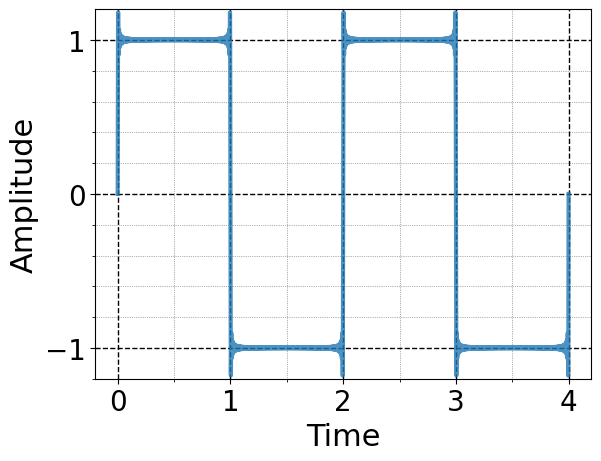}
         \caption{Fourier Reconstruction}
         \label{fig:fourier-recon}
    \end{subfigure}
    \hfill
    \begin{subfigure}[b]{0.45\columnwidth}
         \centering
         \includegraphics[width=\columnwidth]{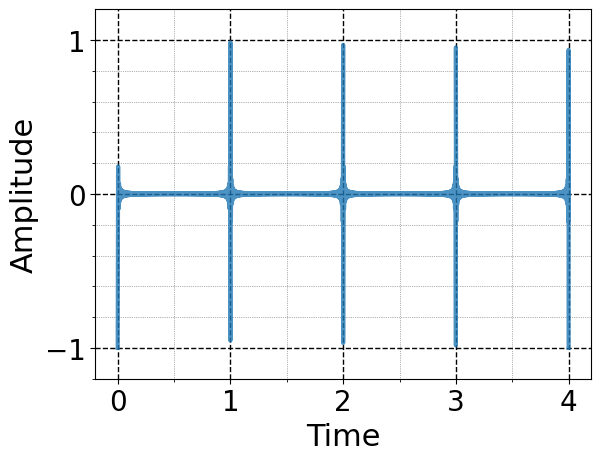}
         \caption{Fourier Error Function}
         \label{fig:fourier-error}
    \end{subfigure}
    \\
    \centering
    \begin{subfigure}[b]{0.45\columnwidth}
         \centering
         \includegraphics[width=\columnwidth]{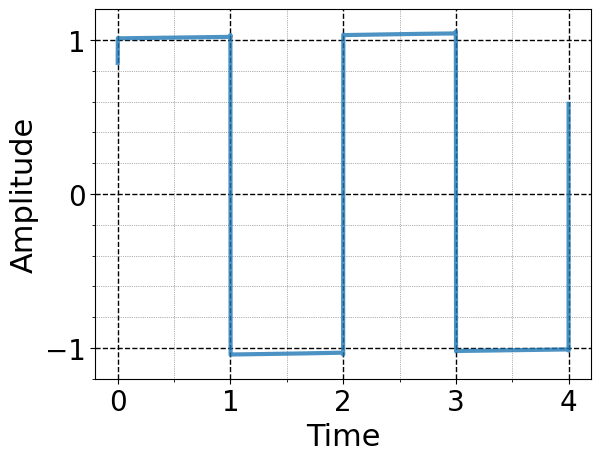}
         \caption{Wavelet Reconstruction}
         \label{fig:wavelet-recon}
    \end{subfigure}
    \hfill
    \begin{subfigure}[b]{0.45\columnwidth}
         \centering
         \includegraphics[width=\columnwidth]{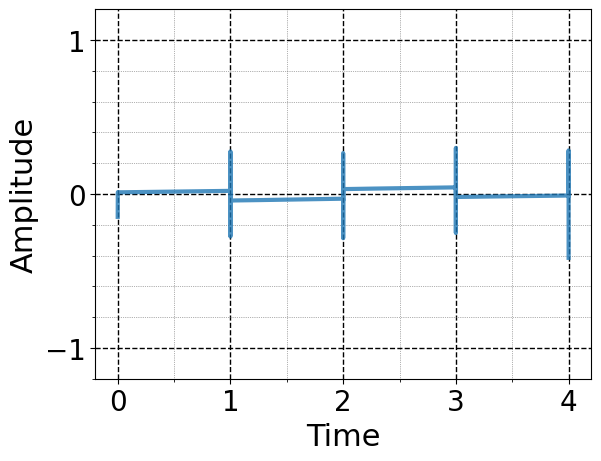}
         \caption{Wavelet Error Function}
         \label{fig:wavelet-error}
    \end{subfigure}
    \caption{The visualization of the reconstructed square wave and the associated error function with different decomposition basis. It can be observed that the wavelet basis can reconstruct the signal with a smaller error regarding the step transient.
    }
    \label{fig:error}
\end{figure}

\begin{figure*}[htp]
    \centering
    \includegraphics[width=0.8\textwidth]{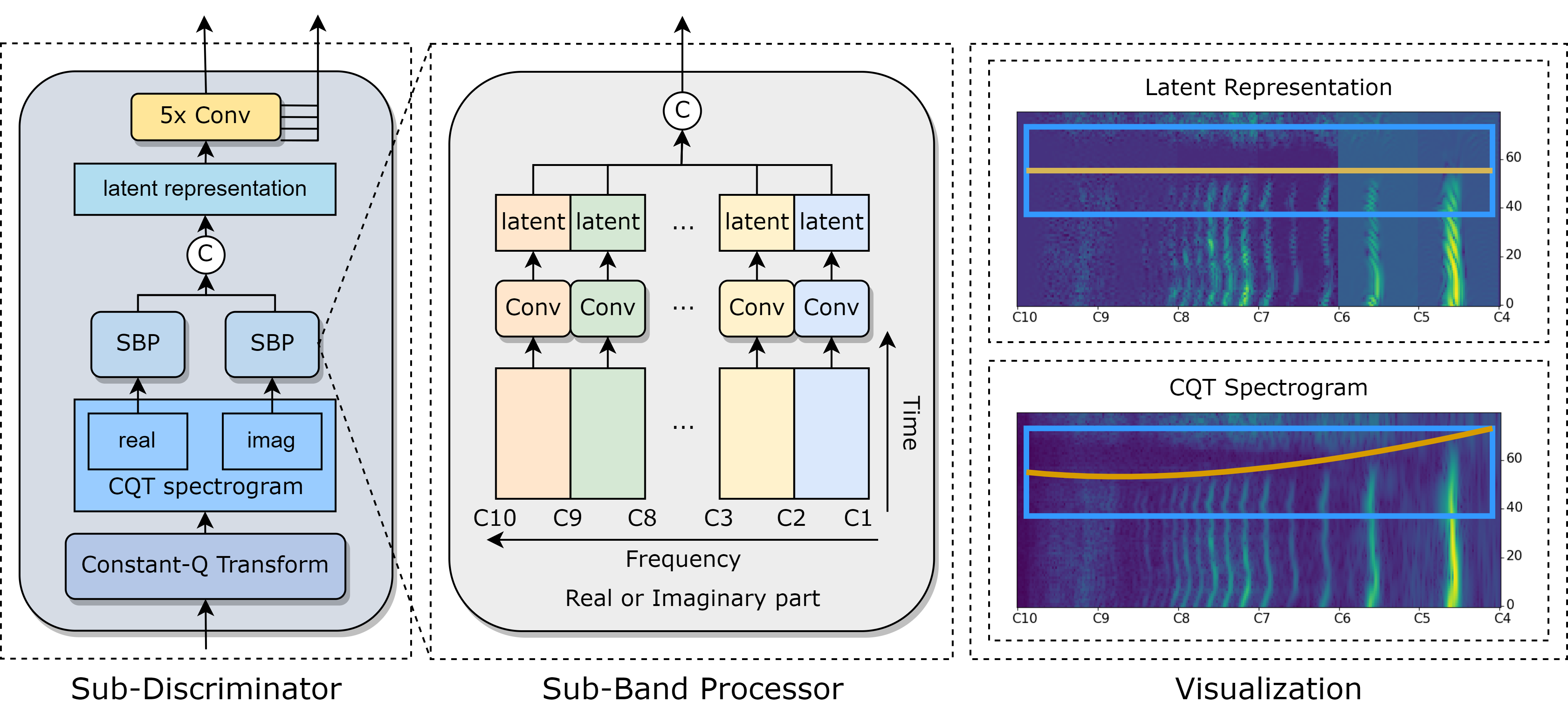}
    \caption{Architecture of the Sub-Discriminator in MS-SB-CQT Discriminator. Operator ``C'' denotes for concatenation. SBP means our proposed Sub-Band Processor module. It can be observed that the desynchronized CQT Spectrogram (bottom-right) has been synchronized (upper-right) after SBP.}
    \label{fig:model_cqt}
\end{figure*}

\subsection{Continuous Wavelet Transform (CWT)}

Continuous Wavelet Transform (CWT)~\cite{CWT1992} converts a time-domain signal $x$ into its TFR with shifting and scaling, as:
\begin{equation}
\begin{split}
\label{eq:cwt}
    \boldsymbol{X} ^ {CWT} (k, n) = \frac{1}{|a _ {k}| ^ \frac{1}{2}}\sum_{j = 1 }^{N} x (j) \psi ^ \ast (\frac{j - n}{a _ {k}}),
\end{split}    
\end{equation}
where $a _ {k}$ is the scaling factor, $\psi (n)$ is the mother wavelet, and $\frac{1}{|a _ {k}| ^ \frac{1}{2}} \psi (\frac{j - n}{a _ {k}})$ is the scaled child wavelet.

By taking the FT between the mother wavelet and the child wavelet, the bandwidth $\Delta f _ {k}$ can be obtained as:
\begin{equation}
\begin{split}
\label{eq:wavelet-bandwidth}
    \Delta f _ {k} = \frac{\Delta f}{a _ {k}},
\end{split}
\end{equation}
where $\Delta f$ is the bandwidth of the mother wavelet. The relationship between the scaling factor $a _ {k}$ and the center frequency $f _ {k}$ is defined as:
\begin{equation}
\label{eq:wavelet-center-freq}
    f _ {k} = \frac{f _ {s}}{a _ {k}},
\end{equation}
thus also obtaining a constant Q-factor:
\begin{equation}
\begin{split}
\label{eq:wavelet-quality-factor}
    Q _ {k} \overset{\textit{ref.}}{=} \frac{f _ {k}}{\Delta f _ {k}} = \frac{f _ {s}}{\Delta f}
\end{split}
\end{equation}

Compared with STFT, CWT tackles the issue brought by the constant TF resolution and the fixed decomposition basis (Table~\ref{tab:three-transforms}), bringing the following benefits:

\begin{itemize}
    \item \textbf{Dynamic TF resolution:} CWT owns a constant Q-factor (Eq.~\ref{eq:wavelet-quality-factor}), which brings a dynamic TF resolution (Fig.~\ref{fig:resolution}).
    \item \textbf{More diverse TFRs:} In CWT, the decomposition basis $\psi (n)$ is a variable. Hence, it is possible to decompose a signal with different $\psi(n)$s for more diverse TFRs. Specifically, to model phase information, this study employs the Complex Morlet Wavelet (CMOR)~\cite{morlet} and the Complex Gaussian Wavelet Family (CGAU)~\cite{gaussian}.
    \item \textbf{Enhanced capability in modeling of short-time transients:} The energy-centralized characteristic of wavelet basis also guarantees CWT a better modeling ability in short-time transients. Fig.~\ref{fig:error} compares the decomposition basis in STFT and CWT when modeling a square wave. Wavelet basis can achieve a reconstructed signal with the same number of components but a significantly smaller error in the places where the "step transients" occur due to its "harsher shape," hence showing a better modeling ability in short-time transients.
\end{itemize}


\section{Methodology}

As discussed in Section~\ref{sec:related-work}, compared with STFT, CQT and CWT hold better modeling ability in expressive speech and singing voice. To utilize them, we propose the MS-SB-CQT Discriminator and MS-TC-CWT Discriminator and introduce a joint training strategy to use discriminators based on STFT, CQT, and CWT in the same framework. Specifically, we modify Multi-Resolution Processing for CQT and CWT and introduce Multi-Basis Processing for CWT in Section~\ref{sec:multi-scale}, propose a Sub-Band Processor for CQT in Section~\ref{sec:sbp}, propose a Temporal Compressor for CWT in Section~\ref{sec:tc}, and elaborate the integration with the generator in Section~\ref{sec:integration}.







\subsection{Multi-Scale Processing for CQT and CWT Discriminators}
\label{sec:multi-scale}

Multi-Scale Processing applies Sub-Discriminators operating on diverse TFRs to reduce the bias brought by the fixed pattern of a specific TF analysis technique, which has been widely used~\cite{HiFiGAN, UniversalMelGAN, encodec, FreGAN, SingGAN}. Motivated by the concept of Multi-Scale Processing, we apply Multi-Resolution Processing to both CQT- and CWT-based discriminators and propose Multi-Basis Processing for CWT-based discriminators only.

\subsubsection{Multi-Resolution Processing}

Multi-Resolution Processing utilizes TFRs with different TF resolution distributions for discriminating to alleviate the TF resolution trade-off bias due to the Uncertainty Principle~\cite{uncertainty}.

For the CQT-based discriminator, Given Eq.~(\ref{eq:cqt-Qfactor}) and (\ref{eq:center_freq}):
\begin{equation}
\label{eq:bandwidth-final}
\begin{split}
    \Delta f _ {k} = \frac{f _ {1} \cdot 2 ^ {\frac{k - 1}{B}}}{(2 ^ {\frac{1}{B}} - 1) ^ {-1}}
\end{split}
\end{equation}

Thus, the bandwidth $\Delta f_k$, which determines the TF resolution trade-off of the $k$-th frequency bin, depends on the number of bins per octave $B$. We use different $B$s to obtain spectrograms with different TF resolution distributions.

For the CWT-based discriminator, as illustrated in Eq.~\ref{eq:wavelet-bandwidth}, the bandwidth $\Delta f _ {k}$ is dependent on the scaling factor $a _ {k}$. Thus, we utilize different scaling factor series to obtain spectrograms with different TF resolution distributions. 

\subsubsection{Multi-Basis Processing}

We propose the Multi-Basis Processing as an extra boost for Multi-Scale Processing on the CWT-base discriminator since CWT has a flexible decomposition basis $\psi (n)$. Specifically, the wavelet basis does not conform to the physical property that the sound comprises a series of sinusoidal components; thus, decomposing signals into such a domain will typically bring biases. To alleviate that, we utilize different wavelet bases to obtain CWT spectrograms in different decomposition domains. Since preliminary experiments verify an independent role between scaling factor $a$ and decomposition basis $\psi (n)$, we implement Multi-Basis Processing in parallel with Multi-Scale Processing for saving up memory, employing the CMOR, the 1-st and 8-st derivatives of CGAU as three distinct complex wavelet bases. 

\subsection{Sub-Band Processor for CQT Discriminator}
\label{sec:sbp}

As two sides of a coin, although the dynamic $\Delta f_k$ brings flexible TF resolution, it also brings the unfixed $N_k$, which can be obtained from Eq.~(\ref{eq:Qfactor-def}), (\ref{eq:bandwidth-def}), and (\ref{eq:cqt-Qfactor}):

\begin{equation}
\label{eq:bandwidth}
    N_k = \frac{f_s}{f _ {k}} \cdot (2 ^ {\frac{1}{B}} - 1) ^ {-1}
\end{equation}

Thus, in a fixed time step $t$, the kernels $a_{k} (t)$ in different frequency bins will convolute different amounts of sample points of the original signal as shown in Eq.~\ref{eq:cqt}. This means the convolutional kernels are not temporally synchronized~\cite{CQT2010} and will cause artifacts in the resulting spectrogram visualized in the bottom right of Fig.~\ref{fig:model_cqt}. 

To alleviate this problem, \cite{CQT2010} designs a series of temporally synchronized kernels within each octave. This algorithm has also been widely used in toolkits like librosa~\cite{librosa} and nnAudio~\cite{nnaudio}. However, such an algorithm only makes the $a_{k} (t)$ of \textit{intra-octave} temporally synchronized but leaves those of \textit{inter-octave} unsolved. Training our neural vocoder using features with such a bias will cause a burden to the adversarial training process and harm the resulting synthesis quality.

Based on that, we utilize the philosophy of representation learning and design the Sub-Band Processor (SBP) module to address this problem further (Fig.~\ref{fig:model_cqt}). In particular, the real or imaginary part of a CQT spectrogram will first be split into sub-bands according to different octaves. Then, each band will be sent to a layer to get its representation. Finally, we concatenate the representations from all bands to obtain the latent representation of the CQT spectrogram. In the upper right of Fig.~\ref{fig:model_cqt}, it can be observed that the SBP module successfully learns the desired representation that is temporally synchronized among all the frequency bins.

\begin{figure*}[htp]
    \centering
    \includegraphics[width=0.8\textwidth]{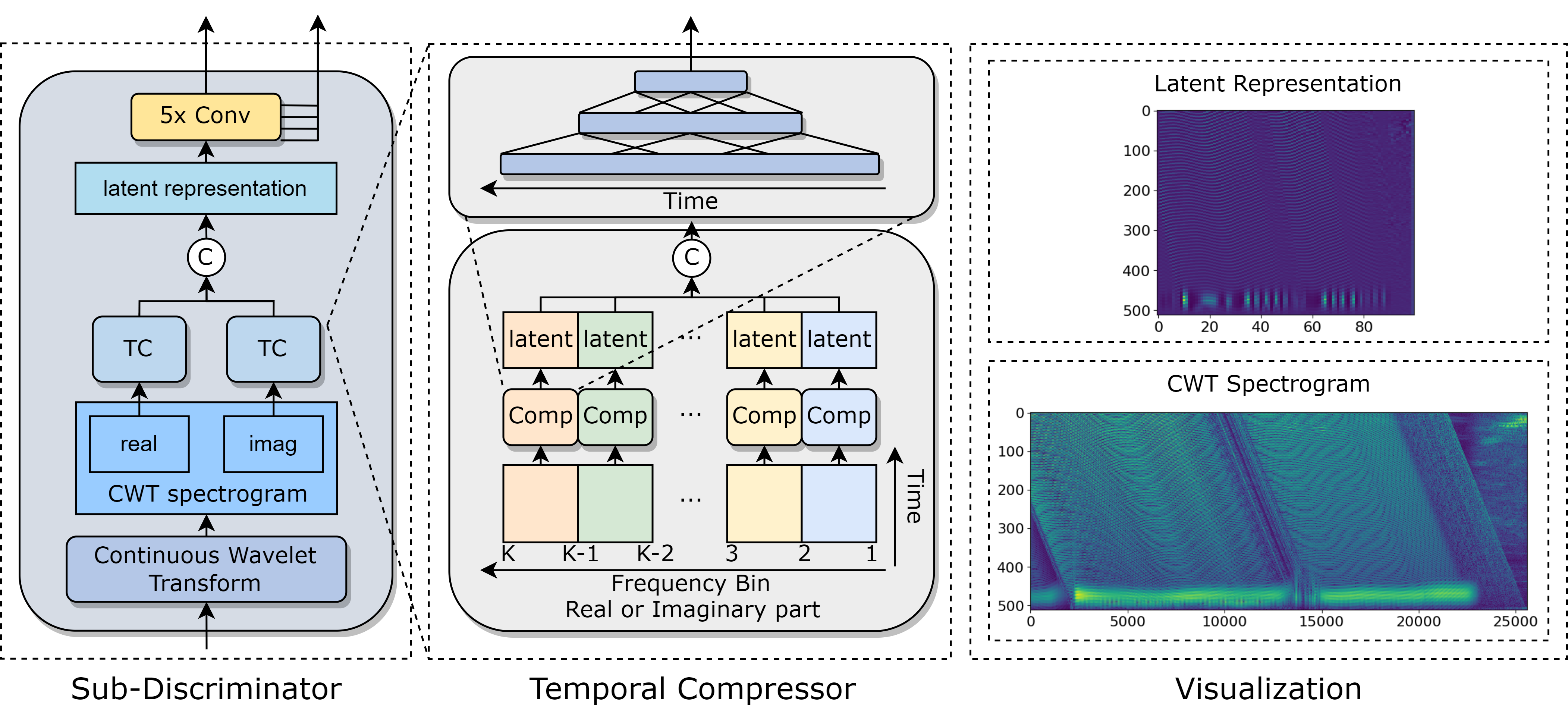}
    \caption{Architecture of the Sub-Discriminator in MS-TC-CWT Discriminator. Operator ``C'' denotes for concatenation. TC means our proposed Temporal Compressor module. Comp is a series of temporal-overlapped convolution layers. $K$ is the total number of frequency bins. It can be observed that the CWT Spectrogram (bottom-right) can be compressed while maintaining the overall energy distribution over different frequency bins (upper-right). 
    }
    \label{fig:model_cwt}
\end{figure*}

\subsection{Temporal Compressor for CWT Discriminator}
\label{sec:tc}

Although the variable wavelet basis $\psi (n)$ brings better TFR diversity and modeling ability in short-time transients, it also has drawbacks. Specifically, the characteristics of wavelet basis bring the requirement of "unit hop length"~\cite{pywavelet} for alleviating information loss and maintaining good invertibility. Thus, the resulting CWT spectrogram from an audio signal of shape (1, T) will have a shape of (Scales, T). Such a large spectrogram will be incompatible with deep learning applications with limited GPU memory.


To reduce the memory complexity, instead of directly utilizing the raw CWT spectrogram, we use the learnable compressed latent representation obtained via our designed Temporal Compressor (TC) module (Fig.~\ref{fig:model_cwt}). Specifically, the real or imaginary part of the CWT spectrogram will first be sent scale-wise to a series of Conv2D layers, which use temporal-overlapped convolutional windows to maintain a good continuity between frames, to get its temporal-compressed representation. Then, we concatenate the representations from all scales to obtain the resulting latent representation of the CWT spectrogram. In the upper right of Fig.~\ref{fig:model_cwt}, it can be observed that our proposed TC module successfully compressed the spectrogram while maintaining the overall energy distribution.

\subsection{Joint Training and Integration with Generators}
\label{sec:integration}

Our proposed discriminators can be easily integrated with any GAN-based vocoders without interfering with the inference stage. We take HiFi-GAN~\cite{HiFiGAN} as an example. HiFi-GAN has a generator $G$ and multiple discriminators $D _ {m}$. The generation loss $\mathcal{L} _ {G}$, and discrimination loss $\mathcal{L} _ {D}$ are as follows:
\begin{equation}
\begin{split}
\mathcal{L} _ {G} = \sum_{m = 1} ^ {M} [\mathcal{L} _ {adv} (G; D _ {m}) + 2 \mathcal{L} _ {fm} (G; D _ {m})] + 45 \mathcal{L} _ {mel},
\end{split}
\end{equation}
\begin{equation}
\begin{split}
\mathcal{L} _ {D} = \sum_{m = 1} ^ {M} [\mathcal{L} _ {adv} (D _ {m}; G),
\end{split}
\end{equation}
where M is the number of discriminators, $D _ {m}$ denotes the m-th discriminator, $\mathcal{L} _ {adv}$ is the adversarial GAN loss, $\mathcal{L} _ {fm}$ is the feature matching loss, and $\mathcal{L} _ {mel}$ is the mel spectrogram reconstruction loss. 

Among these losses, only $\mathcal{L} _ {fm}$ and $\mathcal{L} _ {adv}$ are related to our discriminator. Suppose we want to integrate a new discriminator $D _ {\text{new}}$ in the training process, just adding $\mathcal{L} _ {adv} (G; D _ {\text{new}}) + 2 \mathcal{L} _ {fm} (G; D _ {\text{new}})$ to $\mathcal{L} _ {G}$ and $\mathcal{L} _ {adv} (D _ {\text{new}}; G)$ to $\mathcal{L} _ {D}$ can finish the integration process.

\section{Experiments}
\label{sec4}

We conduct experiments to explore the following questions:
\begin{itemize}
    \item \textbf{EQ1}: Effectiveness of the proposed MS-SB-CQT and MS-TC-CWT Discriminators.
    \item \textbf{EQ2}: Effectiveness of using MS-SB-CQT, MS-TC-CWT and MS-STFT Discriminators jointly.
    \item \textbf{EQ3}: Generalization of the proposed discriminators to various GAN-based vocoders. 
    \item \textbf{EQ4}: Effectiveness of the proposed Sub-Band Processor module and the Multi-Basis Processing technique.
    \item \textbf{EQ5}: What do the latent representations in MS-SB-CQT and MS-TC-CWT Discriminators exactly learn? 
\end{itemize}



\subsection{Experiment Setup}

\subsubsection{Datasets}


The experimental datasets for Section~\ref{sec:effctiveness} and Section~\ref{sec:generalization} contain both speech and singing voices. For the singing voice, we adopt M4Singer~\cite{M4Singer}, PJS~\cite{PJS}, and one internal dataset. We randomly sample 352 utterances from the three datasets to evaluate \textbf{\textit{seen}} singers and leave the remaining for training (39 hours). 445 samples from Opencpop~\cite{Opencpop}, PopCS~\cite{POPCS}, OpenSinger~\cite{OpenSinger}, and CSD~\cite{csd} are chosen to evaluate \textbf{\textit{unseen}} singers. For the speech, we use the train-clean-100 from LibriTTS~\cite{LibriTTS} and LJSpeech~\cite{ljspeech}. We randomly sample 2,316 utterances from the two datasets to evaluate \textbf{\textit{seen}} speakers and leave the remaining for training (about 75 hours). 3,054 utterances from VCTK~\cite{vctk} are chosen to evaluate \textbf{\textit{unseen}} speakers. As for Section~\ref{sec:ablation}, we adopt Opencpop~\cite{Opencpop}. We randomly selected 221 utterances for evaluation and the remaining for training (about 5 hours). The detail of each dataset is listed in Table~\ref{tab:datasets}.

\begin{table}[t]
\centering
\caption{Statistics of the datasets used for training and evaluating the vocoders. LibriTTS, LJSpeech, and VCTK are three speech datasets, while the others are singing voice datasets. EN, CN, KR, and JP means English, Chinese, Korean, and Japanese individually.}\label{tab:dataset-statistics}

\label{tab:datasets}
\resizebox{\columnwidth}{!}{%
\small
\begin{tabular}{lcrrr}
\toprule
\textbf{Dataset} & \textbf{Language}  & \textbf{\#Hours} & \textbf{\#Utterances} & \textbf{\#Speakers} \\ \midrule
LJSpeech~\cite{ljspeech}  & EN & 23.9 & 13,100 & 1 \\
LibriTTS~\cite{LibriTTS} & EN & 585.8 & 375,086 & 2,456 \\
VCTK~\cite{vctk} & EN     & 82.9  &  33,971  & 109  \\ 
\midrule
Internal Dataset & CN & 5.1 & 3,561 & 1 \\
M4Singer~\cite{M4Singer}  & CN & 29.7 &  20,896    & 19  \\
PJS~\cite{PJS} & JP        & 0.4 &  291  & 1  \\
CSD~\cite{csd}  & EN, KR      & 4.1 &  2,864  & 1  \\
OpenSinger~\cite{OpenSinger} & CN & 51.9 &  43,075  & 74  \\
PopCS~\cite{POPCS}  & CN    & 5.9 &  1,651  & 1  \\
Opencpop~\cite{Opencpop} & CN & 5.2 &  3,756  & 1  \\
\bottomrule 
\end{tabular}%
}
\end{table}

\subsubsection{Preprocessing}

We resampled all the training data to 24kHz, excluding the LibriTTS~\cite{LibriTTS} and OpenSinger~\cite{OpenSinger} datasets, which have a sampling rate of 24kHz originally. Each utterance will first be converted to an STFT matrix with an fft size of 1024, hop length of 256, window length of 1024, fmin of 0, and fmax of 12000, which will then be transformed into a mel-spectrogram with 100 mel-filters. The mel-spectrogram is normalized in log-scale with values $\leq$ 0.00001 clipped.

\begin{table*}[t]
\begin{center}
\caption{Analysis-synthesis results of different discriminators when being integrated into HiFi-GAN~\cite{HiFiGAN}. The best and the second best results of every column (except those from Ground Truth) in each domain (speech and singing voice) are \textbf{bold} and \textit{italic}. ``S", ``C" and ``W" represent MS-STFT, MS-SB-CQT and MS-TC-CWT Discriminators respectively. The MOS scores are within 95\% Confidence Interval (CI).}
\label{tab:results-effectiveness}
\resizebox{\textwidth}{!}{
\begin{tabular}{clcccccccccc}

\toprule
\multirow{2}{*}{\textbf{Domain}} & \multirow{2}{*}{\textbf{System}} & \multicolumn{2}{c}{\textbf{PESQ ($\uparrow$)}} & \multicolumn{2}{c}{\textbf{FPC ($\uparrow$)}} & \multicolumn{2}{c}{\textbf{F0RMSE ($\downarrow$)}} & \multicolumn{2}{c}{\textbf{Periodicity ($\downarrow$)}} & \multicolumn{2}{c}{\textbf{MOS ($\uparrow$)}}\\ 
\cmidrule(lr){3-4} \cmidrule(lr){5-6} \cmidrule(lr){7-8} \cmidrule(lr){9-10} \cmidrule(lr){11-12}
& &\textbf{Seen} & \textbf{Unseen} & \textbf{Seen} & \textbf{Unseen} & \textbf{Seen} & \textbf{Unseen} & \textbf{Seen} & \textbf{Unseen} & \textbf{Seen} & \textbf{Unseen} \\
\midrule
\multirow{6}{*}{\makecell{\textbf{Singing}\\ \textbf{voice}}}
& Ground Truth & 4.500 & 4.500 & 1.000 & 1.000 & 0.000 & 0.000 & 0.0000 & 0.0000 & 4.84 $\pm$ 0.07 & 4.86 $\pm$ 0.07 \\ 
\cmidrule(lr){2-12}
 & HiFi-GAN & 2.938 & 2.863 & 0.954 & 0.962 & 56.502 & 60.773 & 0.0675 & 0.0804 & 3.30 $\pm$ 0.17 & 3.61 $\pm$ 0.15 \\
\cmidrule(lr){2-12}
 & HiFi-GAN (+S) & 2.954 & 2.867 & 0.966 & 0.968 & 39.408 & 47.793  & 0.0636 & \textit{0.0734} & 3.44 $\pm$ 0.16 & 3.72 $\pm$ 0.16 \\
 & HiFi-GAN (+C) & \textit{3.031} & \textit{2.947} & \textit{0.968} & 0.971 & \textbf{36.098} & \textit{43.172}  & \textit{0.0620} & 0.0735 & 3.65 $\pm$ 0.15 & 3.83 $\pm$ 0.16 \\
 & HiFi-GAN (+W) & 3.006 & \textbf{2.967} & 0.965 & \textit{0.975} & 42.096 & 47.161  & 0.0644 & 0.0757 & \textit{3.77} $\pm$ \textit{0.14} & \textbf{3.93} $\pm$ \textbf{0.15} \\
\cmidrule(lr){2-12}
 & HiFi-GAN (+S+C+W) & \textbf{3.040} & \textit{2.947} & \textbf{0.972} & \textbf{0.977} & \textit{36.137} & \textbf{42.022} & \textbf{0.0580} & \textbf{0.0717} & \textbf{3.80} $\pm$ \textbf{0.16} & \textit{3.91} $\pm$ \textit{0.17} \\
\midrule
\multirow{6}{*}{\textbf{Speech}}
& Ground Truth & 4.500 & 4.500 & 1.000 & 1.000 & 0.000 & 0.000 & 0.0000 & 0.0000 & 4.83 $\pm$ 0.08 & 4.83 $\pm$ 0.09 \\ 
\cmidrule(lr){2-12}
 & HiFi-GAN & 3.014 & 3.141 & \textit{0.881} & 0.773 & 184.590 & 295.428 & \textbf{0.0062} & 0.0103 & 4.00 $\pm$ 0.18 & 4.07 $\pm$ 0.22 \\
\cmidrule(lr){2-12}
 & HiFi-GAN (+S) & 2.927 & 3.090 & 0.866 & 0.765 & 195.881 & 300.368 & 0.0067 & \textbf{0.0088} & 4.01 $\pm$ 0.18 & 4.10 $\pm$ 0.21 \\
 & HiFi-GAN (+C) & \textit{3.041} & \textit{3.159} & \textit{0.881} & 0.765 & \textit{180.673} & 305.598 & \textbf{0.0062} & 0.0099 & \textit{4.07} $\pm$ \textit{0.18} & 3.98 $\pm$ 0.20 \\
 & HiFi-GAN (+W) & 2.950 & 3.099 & 0.880 & \textit{0.784} & 187.033 & \textit{289.233} & \textit{0.0064} & 0.0102 & \textbf{4.08} $\pm$ \textbf{0.19} & \textit{4.10} $\pm$ \textit{0.19} \\
\cmidrule(lr){2-12}
 & HiFi-GAN (+S+C+W) & \textbf{3.102} & \textbf{3.257} & \textbf{0.882} & \textbf{0.799} & \textbf{178.665} & \textbf{264.935} & 0.0068 & \textit{0.0095} & 4.03 $\pm$ 0.20 & \textbf{4.19} $\pm$ \textbf{0.17} \\
\bottomrule

\end{tabular}
}
\end{center}
\end{table*}

\subsubsection{Training}

All the models are trained using the AdamW~\cite{adamw} optimizer with $\beta _ {1} = 0.8$, $\beta _ {2} = 0.99$, and a initial learning rate of 0.0002. Exponential decay Scheduler is used with a factor $\gamma = 0.999$. All the experiments are conducted on four NVIDIA RTX4090, V100, or A100 GPUs with the batch size of 16 for around 1,500k steps.

\subsubsection{Configurations of Generators and Discriminators}

We use HiFi-GAN~\cite{HiFiGAN}, NSF-HiFiGAN~\cite{POPCS}, BigVGAN~\cite{BigVGAN} and APNet~\cite{APNet} as the experimental generators and Multi-Period Discriminator~\cite{HiFiGAN}, Multi-Scale Discriminator~\cite{HiFiGAN}, Multi-Scale STFT Discriminator~\cite{encodec}, Multi-Scale Sub-Band CQT Discriminator and Multi-Scale Temporal-Compressed CWT Discriminator as the experimental discriminators. The implementation codes are available in Amphion~\cite{amphion}.

The implementation details for the generators are:

\begin{itemize}
        \item \textbf{HiFi-GAN} - The v1 version of the HiFi-GAN~\cite{HiFiGAN}. We reimplemented it using~\footnote{\label{footnote:hifigan}\url{https://github.com/jik876/hifi-gan}} with the same hyperparameters.
        \item \textbf{NSF-HiFiGAN} - The integration of NSF and HiFi-GAN. It is one of the SOTA vocoders for singing voice~\cite{SVCC}. We reimplemented it using~\footnote{\url{https://github.com/MoonInTheRiver/DiffSinger}} with the same hyperparameters.
        \item \textbf{BigVGAN} - The base version of the BigVGAN~\cite{BigVGAN}. It is one of the SOTA vocoders for speech synthesis. We reimplemented it using~\footnote{\label{footnote:bigvgan}\url{https://github.com/NVIDIA/BigVGAN}} with the same hyperparameters.
        \item \textbf{APNet} - The original version of the APNet~\cite{APNet}. It has a fast inference speed with high synthesis quality. We reimplemented it using~\footnote{\label{footnote:apnet}\url{https://github.com/yangai520/APNet}} with the same hyperparameters.
\end{itemize}

The implementation details for the discriminators are:

\begin{itemize}
    \item \textbf{MSD} - Multi-Scale Discriminator: The discriminator proposed by HiFi-GAN~\cite{HiFiGAN}. We re-implemented it using~\footnoteref{footnote:hifigan} with the same hyperparameters.
    \item \textbf{MPD} - Multi-Period Discriminator: The discriminator proposed by HiFi-GAN~\cite{HiFiGAN}. We re-implemented it using~\footnoteref{footnote:bigvgan}. We made modifications to the number of periods ([2, 3, 5, 7, 11, 17, 23, 37]) to make it more effecitve.
    \item \textbf{MS-STFTD} - Multi-Scale STFT Discriminator: The discriminator proposed by Encodec~\cite{encodec}. We re-implemented it using~\footnote{\url{https://github.com/facebookresearch/encodec}} with the same hyperparameters.
    \item \textbf{MS-SB-CQTD} - Multi-Scale Sub-Band CQT Discriminator: One of the proposed discriminators. The CNN in SBP uses a Conv2D with a kernel size of (3, 9) and a channel of 2 covering both the real and imaginary parts. The CNNs in each Sub-Discriminator consist of a Conv2D with kernel size (3, 8) and 32 output channels, three Conv2Ds with dilation rates of [1, 2, 4] in the time dimension, a stride of 2 over the frequency dimension, and a fixed channel of 32, and a Conv2D with kernel size (3, 3), stride (1, 1) and an output channel of 1. LeakyReLU is used as the activation function after each CNN block in the Sub-Discriminator, and Weight Norm~\cite{weightnorm} is applied to all the CNN blocks. For CQT computation, the global hop length is 256, and the $B$s set for three sub-discriminators are [24, 36, 48]. To cover all the frequency bands given the $f _ {1} = 32.7$, 9 octaves are computed. The waveform will be upsampled from $f _ {s}$ to $2 f _ {s}$ before the computation to avoid the $f _ {max}$ of the top octave hitting the Nyquist Frequency.
    \item \textbf{MS-TC-CWTD} - Multi-Scale Temporal-Compressed CWT Discriminator:
    One of the proposed discriminators. The CNNs in the TC module have the kernel sizes of [(16, 1), (16, 1), (8, 1)], strides of [(8, 1), (8, 1), (4, 1)], paddings of [(8, 0), (8, 0), (4, 0)] and a channel of 2 covering both the real and the imaginary parts, which is equivalent to a window length of 2048 and a hop length of 256. The CNNs in each Sub-Discriminator are the same as the ones in the MS-SB-CQT Discriminator with the same activation function and weight normalization. Regarding CWT computation, the scale factor $a$ equals linearly interpolated numbers of 1 to [512, 256, 128]. 
\end{itemize}

\subsection{Evaluation Metrics}

\subsubsection{Objective Evaluation}

We investigate objective metrics focusing on spectrogram reconstruction, F0 accuracy, and phase distortion. The details are listed below:

\begin{itemize}
    \item \textbf{PESQ} (Perceptual Evaluation of Speech Quality)~\cite{PESQ}: A full-reference algorithm that predicts synthesis quality. We employ the Wide-band raw PESQ score from~\footnote{\url{https://github.com/vBaiCai/python-pesq}}. 
    \item \textbf{F0RMSE} (F0 Root Mean Square Error): The Root Mean Square Error (RMSE) of the log-scale F0 (in cent).
    \item \textbf{FPC} (F0 Pearson Correlation Coefficient): The Pearson Correlation Coefficient of F0 trajectories.
    \item \textbf{Periodicity} distortion~\cite{cargan}: The RMSE of the periodicity, which reflects the glitch artifacts that are caused by phase distortion according to previous works~\cite{cargan, BigVGAN, snakegan}. 
\end{itemize}

\subsubsection{Subjective Evaluation}

We use the Mean Opinion Score (MOS) and ABX Preference Test for subjective evaluation. In each MOS test, a total of 20 utterances (10 in-distribution utterances and 10 out-of-distribution utterances) will be evaluated. Listeners were asked to give a naturalness score between 1 and 5 with an interval of 0.5 for each utterance synthesized by generators trained with different setups as well as the ground truth audio. In each ABX test, a total of 30 utterances (6 comparative pairs from Section~\ref{sec:generalization} and 2 comparative pairs from Section~\ref{sec:ablation}), while each comparative pairs have 6 utterances for evaluating) will be evaluated. Listeners were asked to judge which utterance in each pair had better synthesis quality with the help of the ground truth audio. We invited 20 volunteers who are experienced in the audio generation area to attend the subjective evaluation. Thus, each system in the bellowing MOS test has been graded 200 times, and each pair in the preference test has been graded 120 times. The audio samples are available on our demo page\footnote{\label{footnote:demopage}\url{https://vocodexelysium.github.io/TFR-Discriminators/}}.

\begin{table*}[t]
\begin{center}
\caption{Analysis-synthesis results of our proposed Discriminators when applying on NSF-HiFiGAN~\cite{POPCS}, BigVGAN~\cite{BigVGAN}, and APNet~\cite{APNet} in \textit{singing voice} datasets. The improvements are shown in \textbf{bold}. ``S", ``C" and ``W" represent MS-STFT, MS-SB-CQT and MS-TC-CWT Discriminators respectively.}
\label{tab:results-generalization}
\begin{tabular}{lcccccccccccccccc}

\toprule
 \multirow{2}{*}{\textbf{System}} & \multicolumn{2}{c}{\textbf{PESQ ($\uparrow$)}} & \multicolumn{2}{c}{\textbf{FPC ($\uparrow$)}} & \multicolumn{2}{c}{\textbf{F0RMSE ($\downarrow$)}} & \multicolumn{2}{c}{\textbf{Periodicity ($\downarrow$)}} & \multicolumn{2}{c}{\textbf{Preference ($\uparrow$)}}\\ \cmidrule(lr){2-3} \cmidrule(lr){4-5} \cmidrule(lr){6-7} \cmidrule(lr){8-9} \cmidrule(lr){10-11} \cmidrule(lr){12-13}
  & \textbf{Seen} & \textbf{Unseen} & \textbf{Seen} & \textbf{Unseen} & \textbf{Seen} & \textbf{Unseen} & \textbf{Seen} & \textbf{Unseen} & \textbf{Seen} & \textbf{Unseen} \\
\midrule
Ground Truth & 4.500 & 4.500 & 1.000 & 1.000 & 0.000 & 0.000 & 0.0000 & 0.0000 & / & / \\ 
\midrule
 NSF-HiFiGAN & 3.945 & 3.876 & 0.985 & 0.981 & 27.197 & 34.012 & 0.0377 & 0.0451 & 36.84\% & 39.29\% \\
 NSF-HiFiGAN (+S+C+W) & \textbf{3.980} & \textbf{3.907} & 0.981 & 0.980 & \textbf{25.816} & \textbf{32.100} & \textbf{0.0314} & \textbf{0.0380} & \textbf{63.16}\% & \textbf{60.71}\%  \\
\midrule
 BigVGAN & 3.526 & 3.464 & 0.982 & 0.986 & 22.894 & 26.338 & 0.0772 & 0.0820 & 15.79\% & 4.26\% \\
 BigVGAN (+S+C+W) & \textbf{3.696} & \textbf{3.626} & 0.982 & 0.977 & 28.505 & 37.075 & \textbf{0.0387} & \textbf{0.0449} & \textbf{84.21}\% & \textbf{95.74}\%  \\
\midrule
 APNet & 3.254 & 3.117 & 0.816 & 0.832 & 193.642 & 191.684 & 0.0925 & 0.1086 & 15.79\% & 16.07\% \\
 APNet (+S+C+W) & 3.210 & \textbf{3.119} & \textbf{0.956} & \textbf{0.973} & \textbf{47.365} & \textbf{43.624} & 0.0985 & \textbf{0.1056} & \textbf{84.21}\% & \textbf{83.93}\%  \\
\bottomrule

\end{tabular}
\end{center}
\end{table*}

\subsection{Effectiveness of the Proposed Discriminators and Using Them Jointly (EQ1 \& EQ2)}
\label{sec:effctiveness}

To verify the effectiveness of the proposed discriminators, we take HiFi-GAN with MPD and MSD as the baseline model and enhance it with different discriminators. The results of the analysis-synthesis are illustrated in Table~\ref{tab:results-effectiveness}.

Regarding singing voice, we can observe that: (1) HiFi-GAN (+S), HiFi-GAN (+C), and HiFi-GAN (+W) all outperform HiFi-GAN both subjectively and objectively, confirming the importance of the extra adversarial losses in the frequency domain~\cite{MRD}; (2) Both HiFi-GAN (+C) and HiFi-GAN (+W) outperform the HiFi-GAN (+S) objectively and subjectively, illustrating the effectiveness of utilizing TFRs with dynamic TF resolution; (3) HiFi-GAN (+C) outperforms HiFi-GAN (+W) objectively especially on F0-related metrics, showing the effectiveness of the pitch-aware center frequency distribution. HiFi-GAN (+W) outperforms HiFi-GAN (+C) subjectively, showing the effectiveness of the diverse energy-centered wavelet basis; (4) HiFi-GAN (+S+C+W) outperforms both objectively and subjectively on seen singers while holding better objective results with a similar subjective score on unseen singers, confirming the effectiveness of joint training.

Regarding speech, we can observe that: (1) For seen speakers, HiFi-GAN (+S+C+W) performs best objectively with a similar MOS score with HiFi-GAN (+W) (best) and HiFi-GAN (+C) (second best), illustrating the effectiveness of our proposed methods. (2) For unseen speakers, HiFi-GAN (+S+C+W) performs both objectively and subjectively best, indicating the enhanced generalization ability via utilizing different TFR-based discriminators. (3) Specifically, HiFi-GAN (+C) holds a comparatively higher MOS score in seen speakers but a lower MOS score in unseen speakers. We speculate that although CQT has a dynamic TF resolution which brings a better modeling ability, its generalization ability is degraded due to the incompatibility between the pitch-aware center frequency and speech (non-musical) signal.





\begin{figure}[!ht]
    \centering
    \begin{subfigure}[b]{0.45\linewidth}
         \centering
         \includegraphics[width=\linewidth,height=0.9\linewidth]{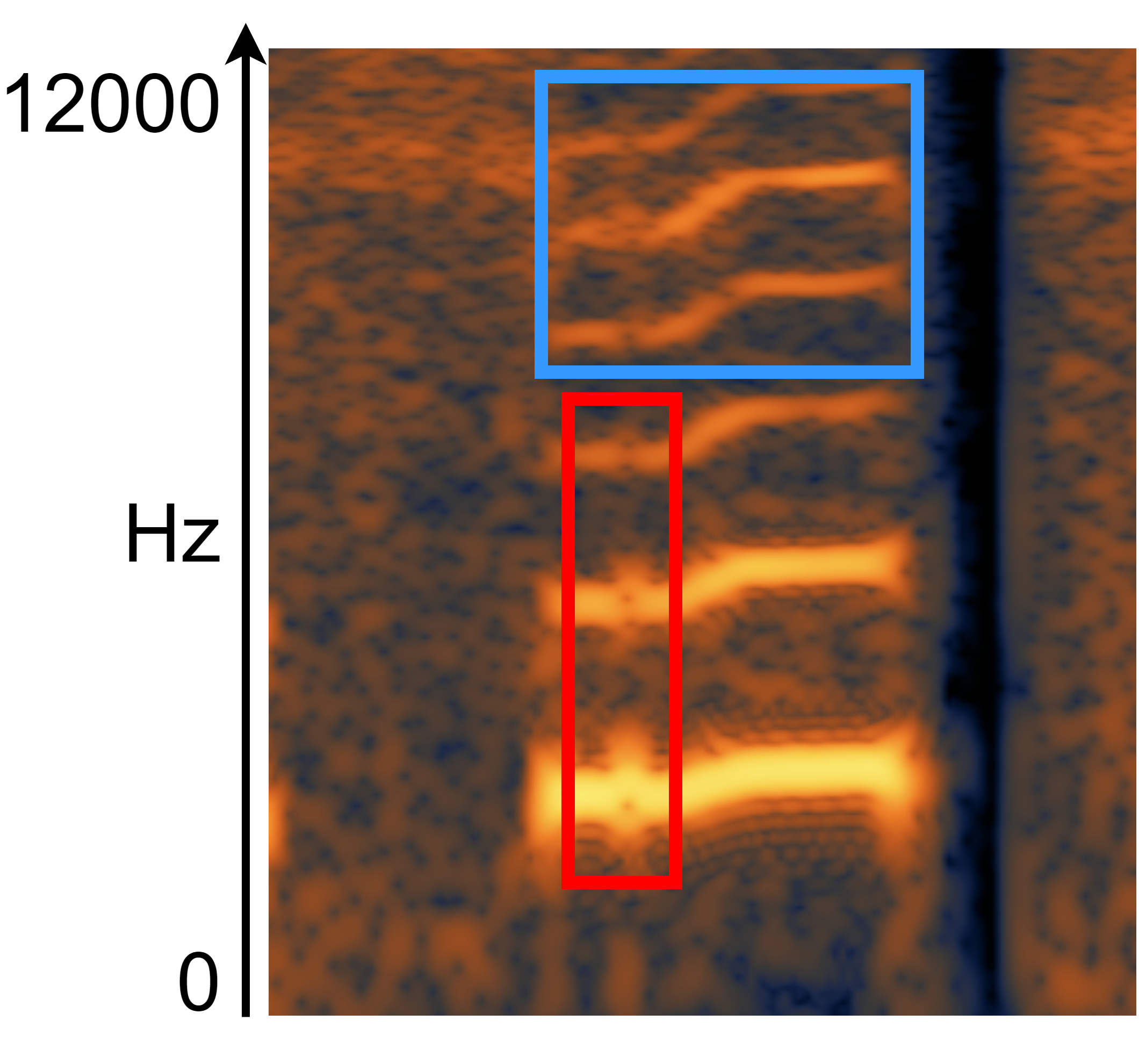}
         \caption{HiFi-GAN (+S)}
         \label{fig:stft}
    \end{subfigure} 
    \hfill
    \begin{subfigure}[b]{0.45\linewidth}
         \centering
         \includegraphics[width=\linewidth,height=0.9\linewidth]{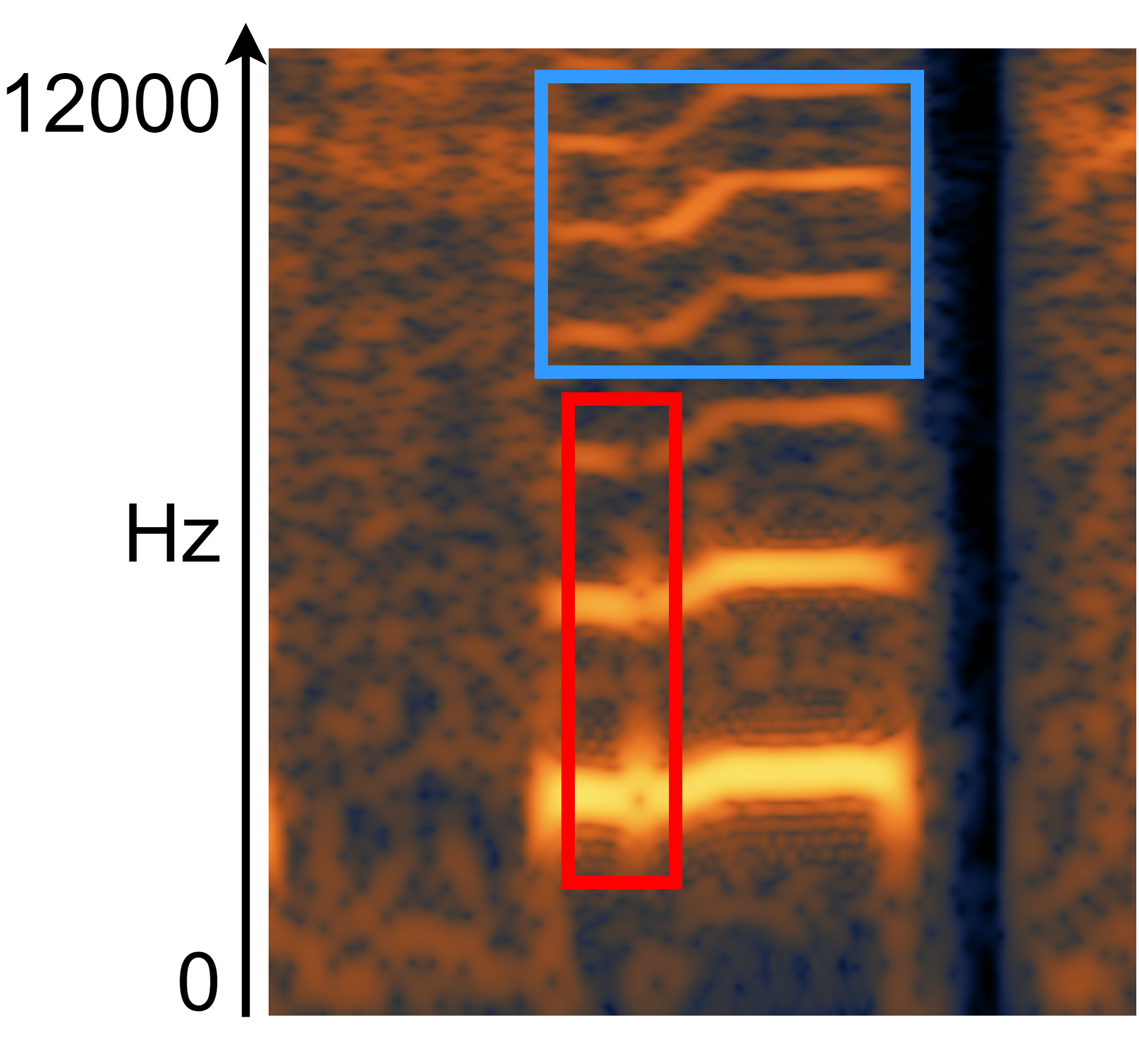}
         \caption{HiFi-GAN (+C)}
         \label{fig:cqt}
    \end{subfigure}
    \newline
    \begin{subfigure}[b]{0.45\linewidth}
         \centering
         \includegraphics[width=\linewidth,height=0.9\linewidth]{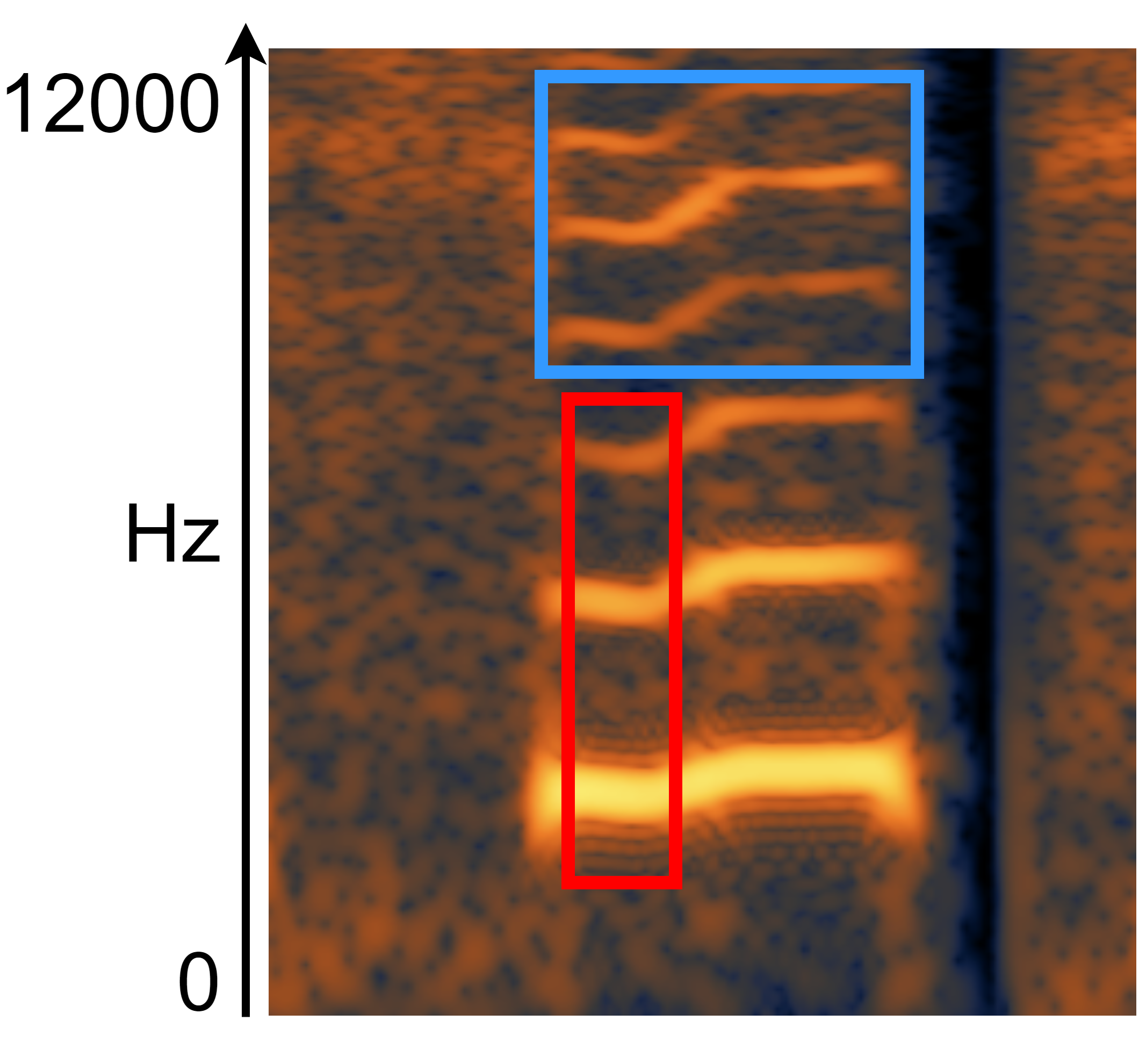}
         \caption{HiFi-GAN (+W)}
         \label{fig:cwt}
    \end{subfigure} 
    \hfill
    \begin{subfigure}[b]{0.45\linewidth}
         \centering
         \includegraphics[width=\linewidth,height=0.9\linewidth]{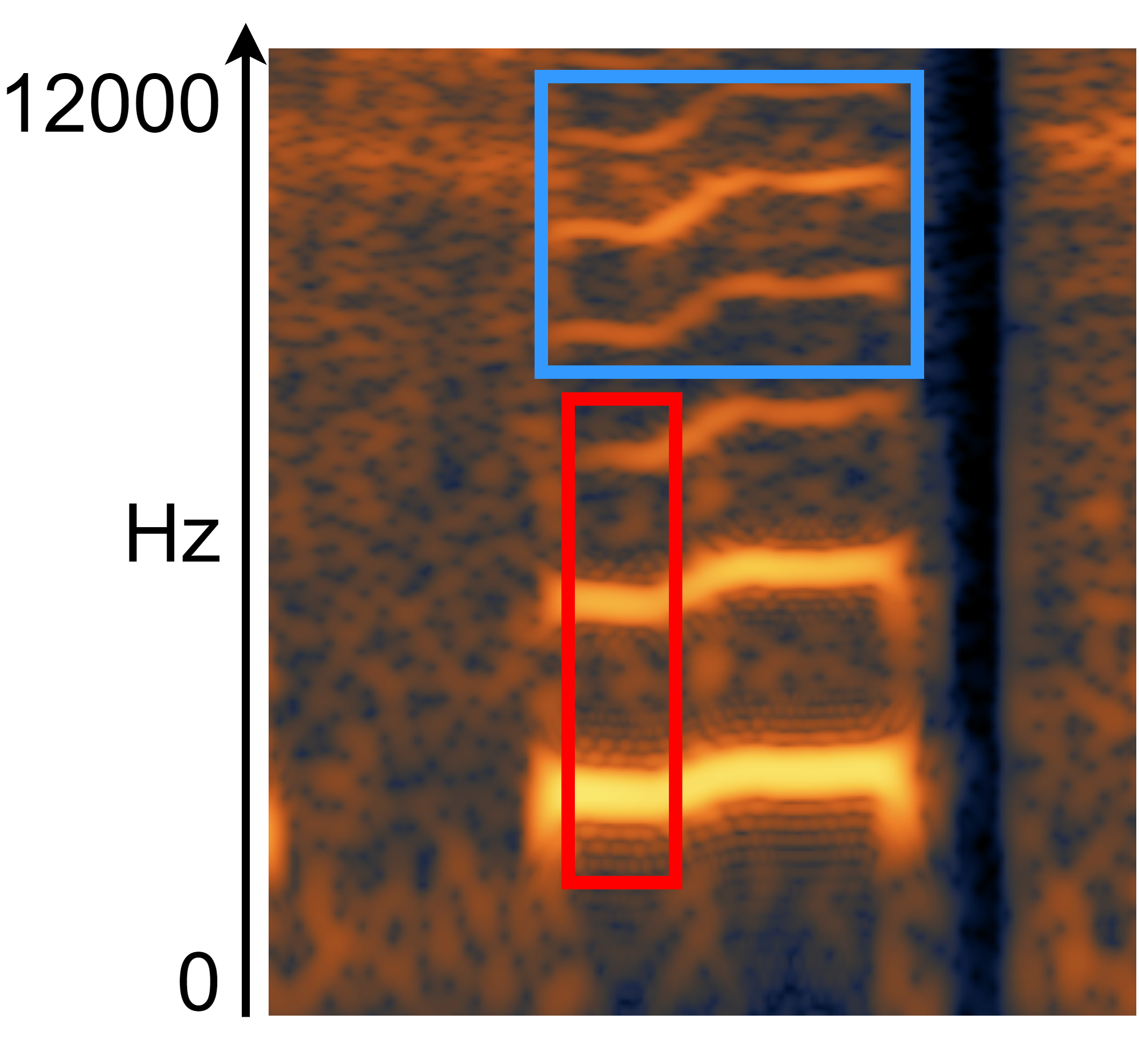}
         \caption{HiFi-GAN (+S+C+W)}
         \label{fig:merge}
    \end{subfigure} 
    \caption{The comparison of mel spectrograms from HiFi-GANs enhanced by different discriminators. ``S", ``C" and ``W" represent MS-STFT, MS-SB-CQT and MS-TC-CWT Discriminators respectively. Integrated with three discriminators, HiFi-GAN could achieve a higher synthesis quality with more accurate harmonic tracking, fundamental frequency reconstruction, and fewer glitches.}
    \label{fig:merge-case-study}
\end{figure}

To further explore the specific benefits of using the STFT-based, CQT-based, and CWT-based discriminators jointly, we conducted a case study as illustrated in Fig.~\ref{fig:merge-case-study}. Notably, in the displayed low-frequency parts, STFT has a better time resolution, while CQT and CWT have a better frequency resolution. It can be observed that: (1) Regarding TF resolution (\textit{upper} rectangles), HiFi-GAN with MS-SB-CQT Discriminator or with MS-TC-CWT Discriminator (Fig.~\ref{fig:cqt}, Fig.~\ref{fig:cwt}) can reconstruct its frequency accurately but "flattened" harmonic component due to the insufficient time resolution, while HiFi-GAN with MS-STFT Discriminator (Fig.~\ref{fig:stft}) can model the transients in the harmonic component but at the expanse of inaccurate pitch information due to the insufficient frequency resolution; (2) Additionally (bottom rectangles), HiFi-GAN with MS-TC-CWT Discriminator (Fig.~\ref{fig:cwt}) can achieve a glitch-free spectrogram, showing its effectiveness in modeling short-time transients. (3) Integrating those three discriminators combines their strengths and thus achieves a better-reconstructed spectrogram (Fig.~\ref{fig:merge}), showing the effectiveness of joint training.

\begin{table*}[t]
\begin{center}
\caption{Analysis-synthesis results of HiFi-GAN enhanced by discriminators with varied modules and processing techniques. The improvements are shown in \textbf{bold}. ``C" and ``W" represent MS-SB-CQT and MS-TC-CWT Discriminators respectively.}
\label{tab:results-ablation}
\begin{tabular}{lccccc}

\toprule
{\textbf{System}} & \textbf{PESQ ($\uparrow$)} & \textbf{FPC ($\uparrow$)} & \textbf{F0RMSE ($\downarrow$)} & \textbf{Periodicity ($\downarrow$)} & \textbf{Preference ($\uparrow$)} \\
 \midrule
Ground Truth & 4.500 & 1.000 & 0.000 & 0.0000 & / \\
 \midrule
HiFi-GAN & 2.960 & 0.972 & 43.139 & 0.0611 & / \\
 \midrule
 \midrule
HiFi-GAN (+W) & 2.880 & \textbf{0.978} & \textbf{40.338} & 0.0705 & \textbf{54.67}\% \\
\quad w/o Multi-Basis Processing technique & 2.898 & 0.969 & 42.306 & 0.0647 & 45.33\% \\
 \midrule
 \midrule
HiFiGAN (+C) & \textbf{2.985} & \textbf{0.985} & \textbf{29.374} & \textbf{0.0634} & \textbf{59.46}\% \\
\quad w/o Sub-Band Processor module& 2.932& 0.963 & 51.162 & 0.0612 & 40.54\% \\

\bottomrule

\end{tabular}
\end{center}
\end{table*}

\subsection{Effectiveness of Proposed Training Strategy (EQ3)}
\label{sec:generalization}

To verify the generalization ability of the complementary role between the STFT-, CQT-, and CWT-based discriminators, we also conduct experiments under NSF-HiFiGAN, BigVGAN, and APNet, as presented in Table~\ref{tab:results-generalization}. 

It is illustrated that: (1) In general, the performance of NSF-HiFiGAN, BigVGAN, and APNet can be improved significantly by jointly training with MS-SB-CQT, MS-TC-CWT, and MS-STFT Discriminators, with both improved objective evaluation scores and subjective preference tests confirming the effectiveness; (2) Regarding NSF-HiFiGAN, although it can synthesis high-fidelity singing voices, it still lacks the modeling abilities of high-frequency band details. Adding adversarial losses based on diverse TFRs alleviates that problem, making synthesized samples closer to the ground truth. Subjective results with a higher preference percentage demonstrate the effectiveness; (3) For BigVGAN, although using the Snake activation function enhanced the generalization ability, it also brings artifacts in phase modeling. Since all MS-STFT, MS-SB-CQT, and MS-TC-CWT Discriminators utilize the phase information, the extra adversarial losses alleviate this problem. Improved subjective scores with significantly higher preference percentages illustrated the effectiveness; (4) As for APNet, although maintaining all the operations in the frame level significantly reduces the memory and computational costs, the difficulty in modeling phase information also brings quality degradation with metallic sound, especially regarding those Aperiodic Parts. Adding MS-STFT, MS-SB-CQT, and MS-TC-CWT Discriminators alleviates that problem, bringing in significantly better FPC and F0-RMSE. We believe this is because the reduction of the metallic sound greatly boosts the performance of the F0-detection algorithm. The subjective evaluation with a higher preference score also confirms its effectiveness. Representative cases regarding these findings can be found on our demo page\footnoteref{footnote:demopage}.


\subsection{Ablation Studies (EQ4)}
\label{sec:ablation}

We propose the Sub-Band Processor module to obtain the temporally synchronized CQT latent representations, and the Multi-Basis Processing technique reduces the bias and thus improves the effectiveness further. We conduct an ablation study of those two methods to verify their necessity. The results of the analysis-synthesis are illustrated in Table~\ref{tab:results-ablation}.

Regarding the CQT-based discriminators, we can see that our proposed MS-SB-CQT Discriminator significantly outperforms the one without the Sub-Band Processor. We speculate this is because the convolutional kernel in the Conv2D layer cannot handle properly the temporal desynchronization in the \textit{inter-octaves} parts of the CQT spectrogram in the initial stage, which would cause a burden and eventually harm the overall audio quality. Regarding the CWT-based discriminators, we can see that our proposed MS-TC-CWT Discriminator performs better than the one that utilizes only one wavelet (we use the CMOR wavelet in the experiment), showing the effectiveness of utilizing different wavelet bases.

\subsection{Analysis on Learned Representation (EQ5)}

\begin{figure}[!ht]
    \centering
    \begin{subfigure}[b]{0.45\linewidth}
         \centering
         \includegraphics[width=\linewidth,height=\linewidth]{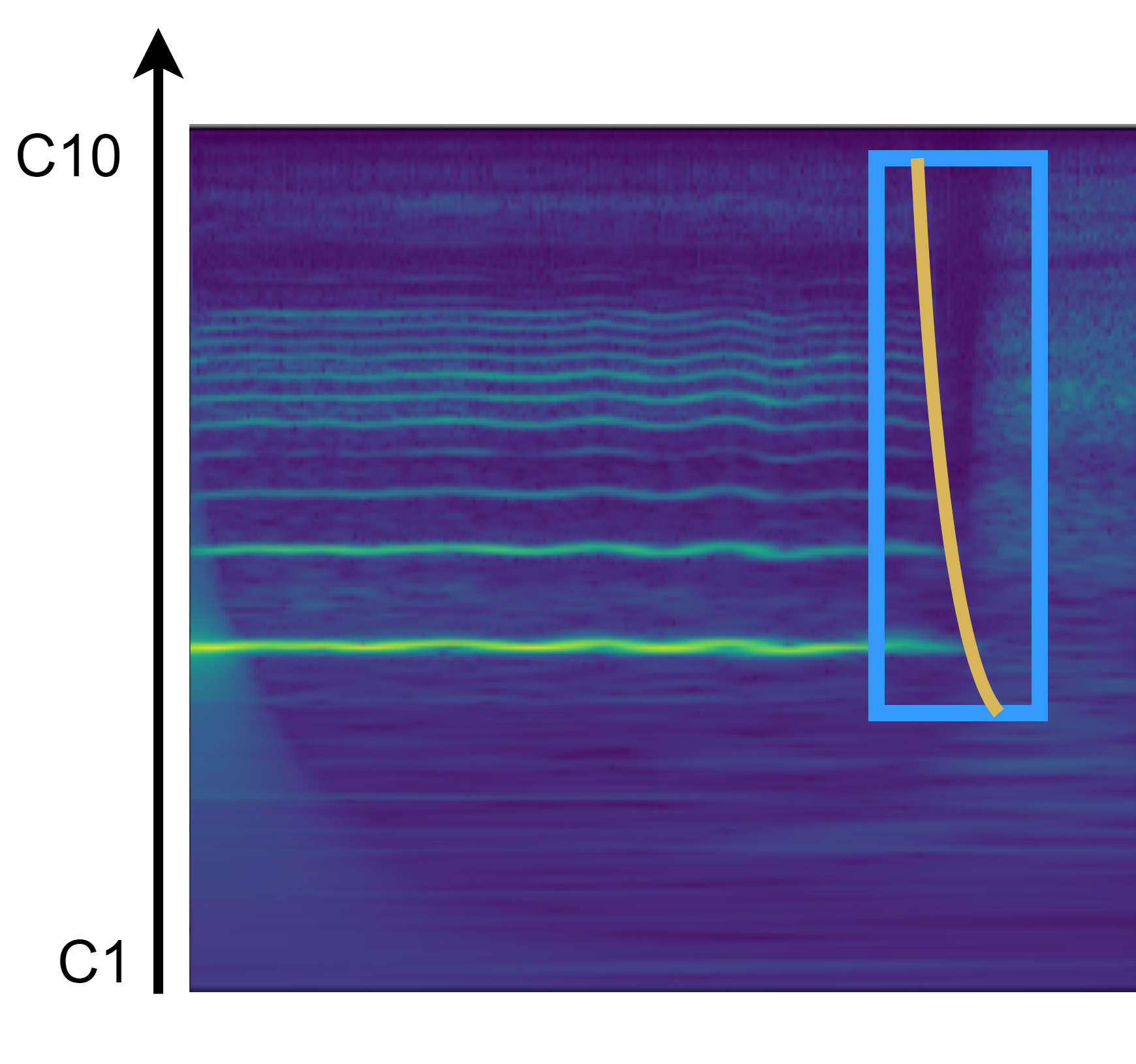}
         \caption{Ground Truth}
         \label{fig:re_gt}
    \end{subfigure}
    \begin{subfigure}[b]{0.45\linewidth}
         \centering
         \includegraphics[width=\linewidth,height=\linewidth]{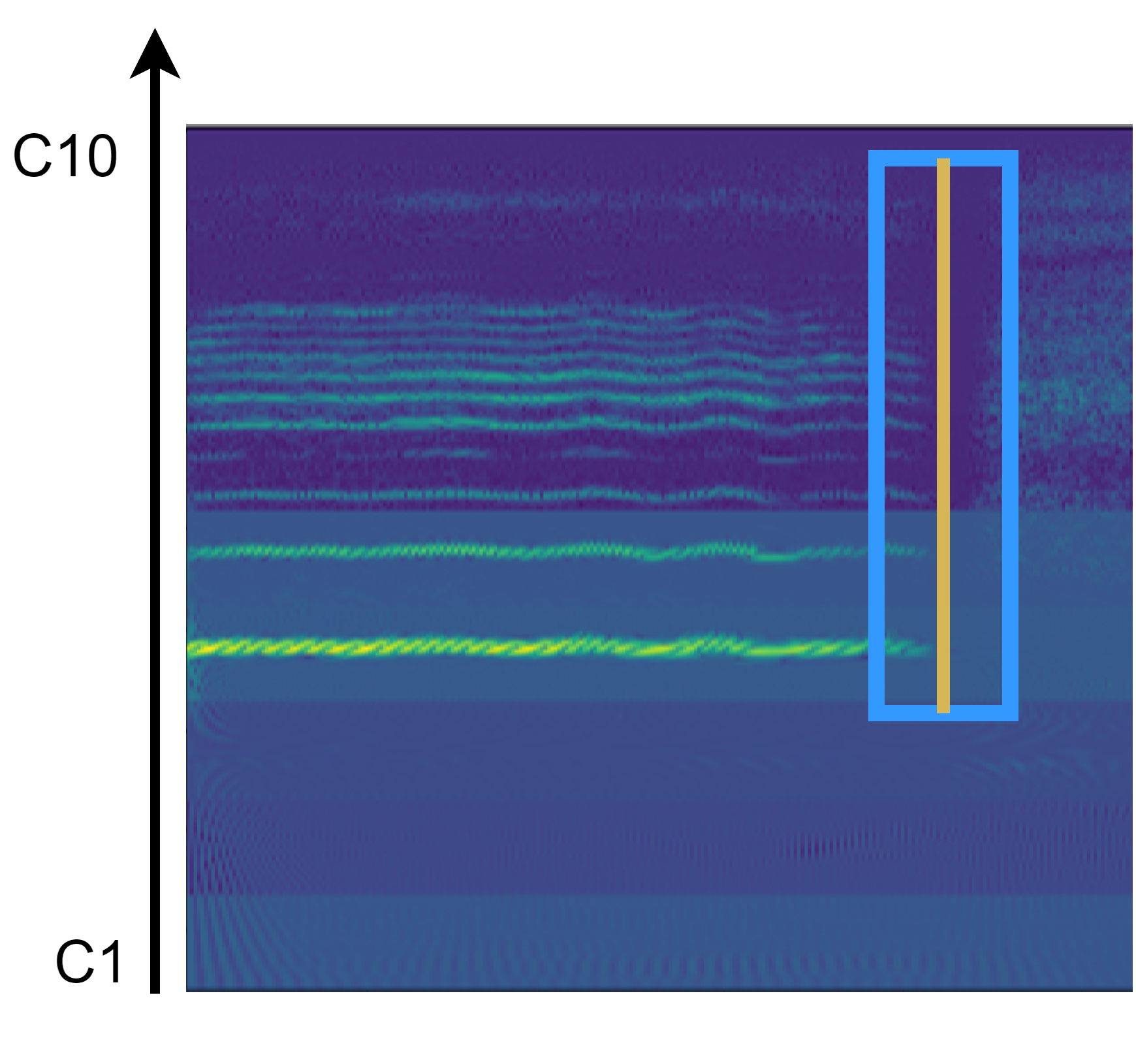}
         \caption{HiFi-GAN}
         \label{fig:re_hifigan}
    \end{subfigure} 
    \begin{subfigure}[b]{0.45\linewidth}
         \centering
         \includegraphics[width=\linewidth,height=\linewidth]{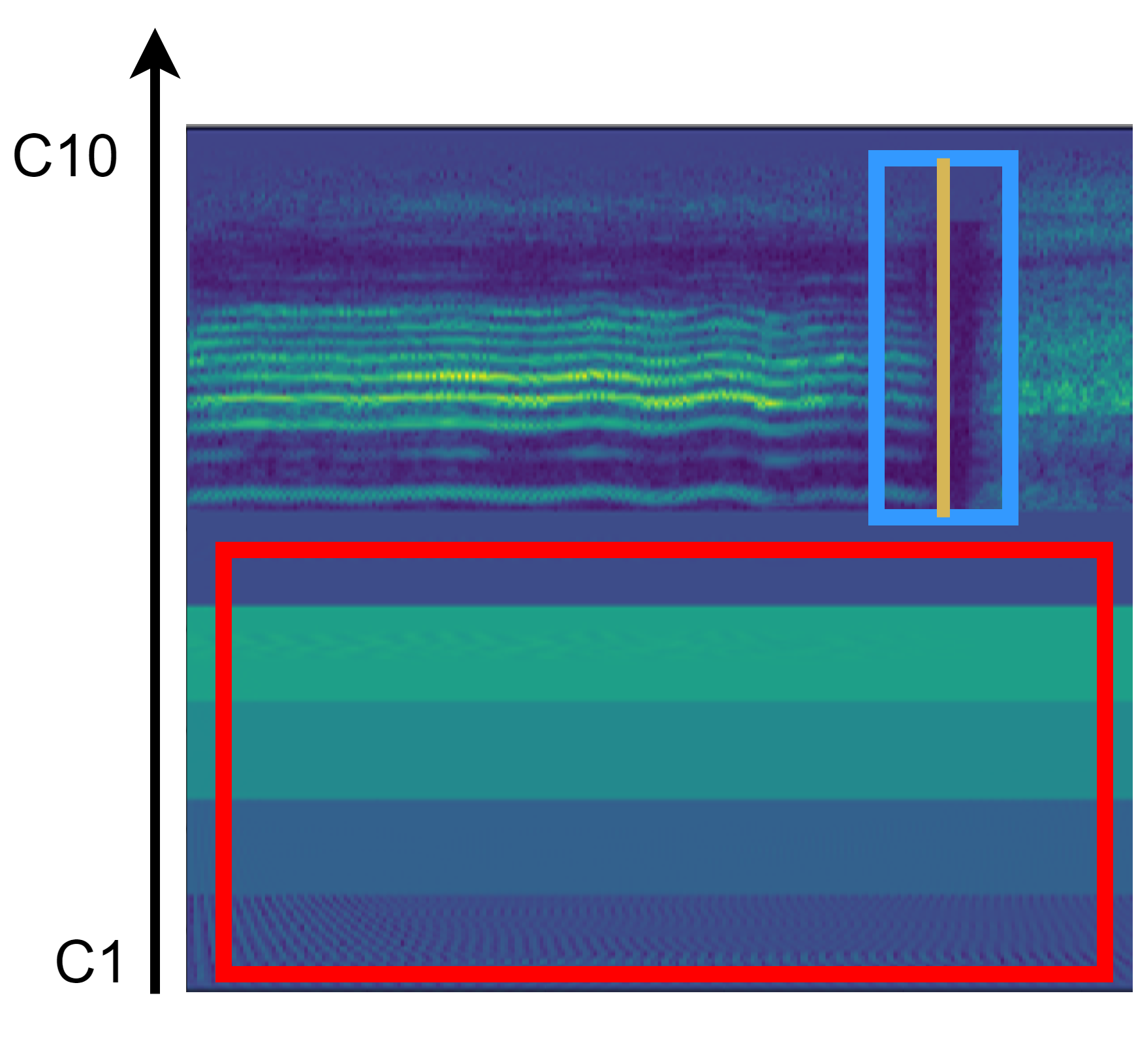}
         \caption{NSF-HiFiGAN}
         \label{fig:re_nsfhifigan}
    \end{subfigure}
    \begin{subfigure}[b]{0.45\linewidth}
         \centering
         \includegraphics[width=\linewidth,height=\linewidth]{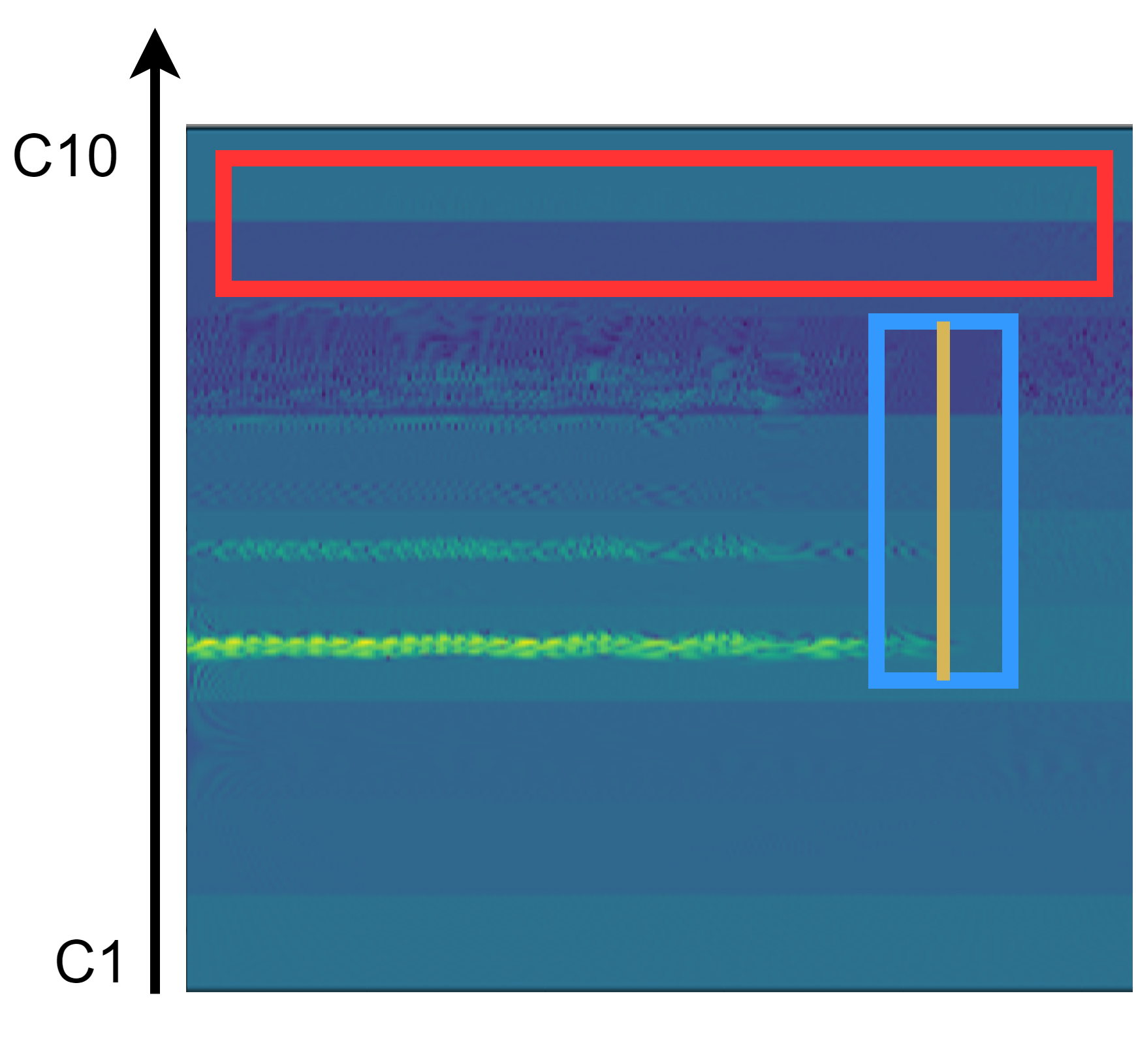}
         \caption{BigVGAN}
         \label{fig:re_bigvgan}
    \end{subfigure} 
    \caption{The comparison of the latent representations from the SBP module trained with different generators. All the SBP modules learn to obtain the temporally synchronized representation, while the ones trained with NSF-HiFiGAN and BigVGAN additionally apply dynamic attention to different frequency bands, masking the frequency band where the generator already holds a high reconstruction quality.}
    \label{fig:representation-study}
\end{figure}

We also conducted a case study on the learned representation in the MS-SB-CQT and MS-TC-CWT Discriminators. Regarding the representation in MS-TC-CWT Discriminator, only compression is learned. Regarding the representation in MS-SB-CQT Discriminator, however, apart from the ability to obtain a temporally synchronized spectrogram, it also learns to apply dynamic attention to different frequency bands (Fig.~\ref{fig:representation-study}).

Notably, among these three vocoders, HiFi-GAN performs averagely across all frequency bands; NSF-HiFiGAN performs better in the low-frequency bands due to its glitch-free ability and accurate pitch modeling but worse in the high-frequency bands due to the aliasing effects; BigVGAN performs better in the high-frequency bands due to its aliasing-free ability but worse in the low-frequency bands due to glitches. It can be observed that: (1) All the SBP modules trained with different generators learned to obtain the temporally synchronized representation (blue rectangle) compared with the ground truth CQT spectrogram; (2) The SBP module trained with NSF-HiFiGAN and BigVGAN learn to mask the low-frequency band and high-frequency band individually(red rectangle). We speculate it is to ignore the frequency band where the generator already has a good synthesis quality and to focus on the frequency bands with a poor reconstruction quality to discriminate whether the audio is generated or not, which indicates the SBP module learns applying dynamic attention to different frequency bands.

\section{Conclusion}
\label{sec5}

This study proposed a Multi-Scale Sub-Band Constant-Q Transform (MS-SB-CQT) Discriminator and a Multi-Scale Temporal-Compressed Continuous Wavelet Transform (MS-TC-CWT) Discriminator for GAN-based Vocoders. Experiments conducted on both speech and singing voices confirm the effectiveness of our proposed methods, with strengths conforming to their design ideas. Moreover, with joint training, the proposed two discriminators can also be complementary with the existing MS-STFT Discriminator to improve the neural vocoder further.



\bibliographystyle{IEEEtran}
\bibliography{ref}

\begin{thebibliography}{10}
\providecommand{\url}[1]{#1}
\csname url@samestyle\endcsname
\providecommand{\newblock}{\relax}
\providecommand{\bibinfo}[2]{#2}
\providecommand{\BIBentrySTDinterwordspacing}{\spaceskip=0pt\relax}
\providecommand{\BIBentryALTinterwordstretchfactor}{4}
\providecommand{\BIBentryALTinterwordspacing}{\spaceskip=\fontdimen2\font plus
\BIBentryALTinterwordstretchfactor\fontdimen3\font minus \fontdimen4\font\relax}
\providecommand{\BIBforeignlanguage}[2]{{%
\expandafter\ifx\csname l@#1\endcsname\relax
\typeout{** WARNING: IEEEtran.bst: No hyphenation pattern has been}%
\typeout{** loaded for the language `#1'. Using the pattern for}%
\typeout{** the default language instead.}%
\else
\language=\csname l@#1\endcsname
\fi
#2}}
\providecommand{\BIBdecl}{\relax}
\BIBdecl

\bibitem{POPCS}
J.~Liu, C.~Li, Y.~Ren, F.~Chen, and Z.~Zhao, ``{DiffSinger: Singing Voice Synthesis via Shallow Diffusion Mechanism},'' in \emph{{AAAI}}, 2022, pp. 11\,020--11\,028.

\bibitem{hiddensinger}
J.~Hwang, S.~Lee, and S.~Lee, ``{HiddenSinger: High-Quality Singing Voice Synthesis via Neural Audio Codec and Latent Diffusion Models},'' \emph{CoRR}, vol. abs/2306.06814, 2023.

\bibitem{fs2}
Y.~Ren, C.~Hu, X.~Tan, T.~Qin, S.~Zhao, Z.~Zhao, and T.~Liu, ``{FastSpeech 2: Fast and High-Quality End-to-End Text to Speech},'' in \emph{ICLR}, 2021.

\bibitem{ns2}
K.~Shen, Z.~Ju, X.~Tan, Y.~Liu, Y.~Leng, L.~He, T.~Qin, S.~Zhao, and J.~Bian, ``{NaturalSpeech 2: Latent Diffusion Models are Natural and Zero-Shot Speech and Singing Synthesizers},'' \emph{CoRR}, vol. abs/2304.09116, 2023.

\bibitem{encodec}
A.~D{\'{e}}fossez, J.~Copet, G.~Synnaeve, and Y.~Adi, ``{High Fidelity Neural Audio Compression},'' \emph{arXiv}, vol. abs/2210.13438, 2022.

\bibitem{funcodec}
Z.~Du, S.~Zhang, K.~Hu, and S.~Zheng, ``{FunCodec: {A} Fundamental, Reproducible and Integrable Open-source Toolkit for Neural Speech Codec},'' \emph{CoRR}, vol. abs/2309.07405, 2023.

\bibitem{straight}
H.~Kawahara, ``{STRAIGHT, exploitation of the other aspect of VOCODER: Perceptually isomorphic decomposition of speech sounds},'' \emph{AST}, pp. 349--353, 2006.

\bibitem{world}
M.~Morise, F.~Yokomori, and K.~Ozawa, ``{WORLD: a vocoder-based high-quality speech synthesis system for real-time applications},'' \emph{IEICE Trans Inf Syst}, vol.~99, no.~7, pp. 1877--1884, 2016.

\bibitem{WaveNet}
A.~van~den Oord, S.~Dieleman, H.~Zen, K.~Simonyan, O.~Vinyals, A.~Graves, N.~Kalchbrenner, A.~W. Senior, and K.~Kavukcuoglu, ``{WaveNet: A Generative Model for Raw Audio},'' in \emph{{SSW}}.\hskip 1em plus 0.5em minus 0.4em\relax {ISCA}, 2016, p. 125.

\bibitem{WaveRNN}
N.~Kalchbrenner, E.~Elsen, K.~Simonyan, S.~Noury, N.~Casagrande, E.~Lockhart, F.~Stimberg, A.~van~den Oord, S.~Dieleman, and K.~Kavukcuoglu, ``{Efficient Neural Audio Synthesis},'' in \emph{{ICML}}, 2018, pp. 2415--2424.

\bibitem{parallelwavenet}
A.~Oord, Y.~Li, I.~Babuschkin, K.~Simonyan, O.~Vinyals, K.~Kavukcuoglu, G.~Driessche, E.~Lockhart, L.~Cobo, F.~Stimberg \emph{et~al.}, ``{Parallel wavenet: Fast high-fidelity speech synthesis},'' in \emph{ICML}, 2018, pp. 3918--3926.

\bibitem{clarinet}
W.~Ping, K.~Peng, and J.~Chen, ``{ClariNet: Parallel Wave Generation in End-to-End Text-to-Speech},'' in \emph{ICLR}, 2019.

\bibitem{WaveFlow}
W.~Ping, K.~Peng, K.~Zhao, and Z.~Song, ``{WaveFlow: A Compact Flow-based Model for Raw Audio},'' in \emph{{ICML}}, vol. 119, 2020, pp. 7706--7716.

\bibitem{WaveGlow}
R.~Prenger, R.~Valle, and B.~Catanzaro, ``{Waveglow: A Flow-based Generative Network for Speech Synthesis},'' in \emph{{ICASSP}}, 2019, pp. 3617--3621.

\bibitem{glotnet}
L.~Juvela, B.~Bollepalli, V.~Tsiaras, and P.~Alku, ``{Glotnet—a raw waveform model for the glottal excitation in statistical parametric speech synthesis},'' \emph{TASLP}, vol.~27, no.~6, pp. 1019--1030, 2019.

\bibitem{lpcnet}
J.-M. Valin and J.~Skoglund, ``{LPCNet: Improving neural speech synthesis through linear prediction},'' in \emph{ICASSP}, 2019, pp. 5891--5895.

\bibitem{DiffWave}
Z.~Kong, W.~Ping, J.~Huang, K.~Zhao, and B.~Catanzaro, ``{DiffWave: A Versatile Diffusion Model for Audio Synthesis},'' in \emph{{ICLR}}, 2021.

\bibitem{fregrad}
T.~D. Nguyen, J.-H. Kim, Y.~Jang, J.~Kim, and J.~S. Chung, ``{FreGrad: Lightweight and Fast Frequency-aware Diffusion Vocoder},'' \emph{arXiv:2401.10032}, 2024.

\bibitem{NSF}
X.~Wang, S.~Takaki, and J.~Yamagishi, ``{Neural Source-filter-based Waveform Model for Statistical Parametric Speech Synthesis},'' in \emph{{ICASSP}}, 2019, pp. 5916--5920.

\bibitem{golf}
C.~Yu and G.~Fazekas, ``{Singing Voice Synthesis Using Differentiable {LPC} and Glottal-Flow-Inspired Wavetables},'' in \emph{ISMIR}, 2023, pp. 667--675.

\bibitem{PWG}
R.~Yamamoto, E.~Song, and J.~Kim, ``{Parallel Wavegan: {A} Fast Waveform Generation Model Based on Generative Adversarial Networks with Multi-Resolution Spectrogram},'' in \emph{{ICASSP}}, 2020, pp. 6199--6203.

\bibitem{MelGAN}
K.~Kumar, R.~Kumar, T.~de~Boissiere, L.~Gestin, W.~Z. Teoh, J.~Sotelo, A.~de~Br{\'{e}}bisson, Y.~Bengio, and A.~C. Courville, ``{MelGAN: Generative Adversarial Networks for Conditional Waveform Synthesis},'' in \emph{NeurIPS}, 2019.

\bibitem{UniversalMelGAN}
W.~Jang, D.~Lim, and J.~Yoon, ``{Universal MelGAN: {A} Robust Neural Vocoder for High-Fidelity Waveform Generation in Multiple Domains},'' \emph{arXiv}, vol. abs/2011.09631, 2020.

\bibitem{HiFiGAN}
J.~Su, Z.~Jin, and A.~Finkelstein, ``{HiFi-GAN: High-Fidelity Denoising and Dereverberation Based on Speech Deep Features in Adversarial Networks},'' in \emph{{INTERSPEECH}}, 2020, pp. 4506--4510.

\bibitem{FreGAN}
J.~Kim, S.~Lee, J.~Lee, and S.~Lee, ``{Fre-GAN: Adversarial Frequency-Consistent Audio Synthesis},'' in \emph{INTERSPEECH}, 2021, pp. 2197--2201.

\bibitem{SingGAN}
R.~Huang, C.~Cui, F.~Chen, Y.~Ren, J.~Liu, Z.~Zhao, B.~Huai, and Z.~Wang, ``{SingGAN: Generative Adversarial Network For High-Fidelity Singing Voice Generation},'' in \emph{{ACM MM}}, 2022, pp. 2525--2535.

\bibitem{BigVGAN}
S.~Lee, W.~Ping, B.~Ginsburg, B.~Catanzaro, and S.~Yoon, ``{BigVGAN: {A} Universal Neural Vocoder with Large-Scale Training},'' in \emph{{ICLR}}, 2023.

\bibitem{bigvsan}
T.~Shibuya, Y.~Takida, and Y.~Mitsufuji, ``{BigVSAN: Enhancing GAN-based Neural Vocoders with Slicing Adversarial Network},'' \emph{CoRR}, vol. abs/2309.02836, 2023.

\bibitem{sawsing}
D.~Wu, W.~Hsiao, F.~Yang, O.~Friedman, W.~Jackson, S.~Bruzenak, Y.~Liu, and Y.~Yang, ``{DDSP-based Singing Vocoders: {A} New Subtractive-based Synthesizer and {A} Comprehensive Evaluation},'' in \emph{ISMIR}, 2022, pp. 76--83.

\bibitem{snake}
Z.~Liu, T.~Hartwig, and M.~Ueda, ``{Neural networks fail to learn periodic functions and how to fix it},'' \emph{NeurIPS}, vol.~33, pp. 1583--1594, 2020.

\bibitem{snakegan}
S.~Li, S.~Liu, L.~Zhang, X.~Li, Y.~Bian, C.~Weng, Z.~Wu, and H.~Meng, ``{SnakeGAN: {A} Universal Vocoder Leveraging {DDSP} Prior Knowledge and Periodic Inductive Bias},'' in \emph{ICME}, 2023, pp. 1703--1708.

\bibitem{MRD}
J.~You, D.~Kim, G.~Nam, G.~Hwang, and G.~Chae, ``{GAN Vocoder: Multi-Resolution Discriminator Is All You Need},'' in \emph{INTERSPEECH}, 2021, pp. 2177--2181.

\bibitem{STFT}
J.~Allen, ``{Short term spectral analysis, synthesis, and modification by discrete Fourier transform},'' \emph{TASLP}, vol.~25, no.~3, pp. 235--238, 1977.

\bibitem{CQT1992}
J.~C. Brown and M.~Puckette, ``{An efficient algorithm for the calculation of a constant Q transform},'' \emph{JASA}, vol.~92, pp. 2698--2701, 1992.

\bibitem{CQT2010}
C.~Sch{\"o}rkhuber and A.~Klapuri, ``{Constant-Q transform toolbox for music processing},'' in \emph{SMCC}, 2010, pp. 3--64.

\bibitem{CWT1992}
O.~Rioul and P.~Duhamel, ``{Fast algorithms for discrete and continuous wavelet transforms},'' \emph{IEEE Trans. Inf. Theory}, vol.~38, no.~2, pp. 569--586, 1992.

\bibitem{CQTICASSP}
Y.~Gu, X.~Zhang, L.~Xue, and Z.~Wu, ``Multi-scale sub-band constant-q transform discriminator for high-fidelity vocoder,'' in \emph{ICASSP 2024-2024 IEEE International Conference on Acoustics, Speech and Signal Processing (ICASSP)}.\hskip 1em plus 0.5em minus 0.4em\relax IEEE, 2024, pp. 10\,616--10\,620.

\bibitem{gibbs}
E.~Hewitt and R.~Hewitt, ``{The Gibbs-Wilbraham phenomenon: An episode in fourier analysis},'' \emph{Arch. Hist. Exact Sci. 21, 129–160}, 1979.

\bibitem{constantQ1}
S.~Seneff, ``{Pitch and spectral analysis of speech based on an auditory synchrony model},'' Ph.D. dissertation, Massachusetts Institute of Technology, Research Laboratory of Electronics, 1985.

\bibitem{constantQ2}
J.~P. Stautner, ``{Analysis and synthesis of music using the auditory transform},'' Ph.D. dissertation, Massachusetts Institute of Technology, 1983.

\bibitem{morlet}
M.~Taner, ``{Joint time/frequency analysis, Q quality factor and dispersion computation using Gabor-Morlet wavelets or the Gabor-Morlet transform},'' \emph{RSI}, pp. 1--5, 1983.

\bibitem{gaussian}
R.~Navarro and A.~Tabernero, ``Gaussian wavelet transform: two alternative fast implementations for images,'' \emph{MSSP}, vol.~2, pp. 421--436, 1991.

\bibitem{uncertainty}
G.~B. Folland and A.~Sitaram, ``The uncertainty principle: a mathematical survey,'' \emph{JFAA}, vol.~3, pp. 207--238, 1997.

\bibitem{librosa}
B.~McFee, C.~Raffel, D.~Liang, D.~P.~W. Ellis, M.~McVicar, E.~Battenberg, and O.~Nieto, ``{librosa: Audio and Music Signal Analysis in Python},'' in \emph{SciPy}, 2015.

\bibitem{nnaudio}
K.~W. Cheuk, H.~Anderson, K.~Agres, and D.~Herremans, ``{nnAudio: An on-the-Fly {GPU} Audio to Spectrogram Conversion Toolbox Using 1D Convolutional Neural Networks},'' \emph{{IEEE} Access}, pp. 161\,981--162\,003, 2020.

\bibitem{pywavelet}
G.~R. Lee, R.~Gommers, F.~Waselewski, K.~Wohlfahrt, and A.~OLeary, ``{PyWavelets: A Python package for wavelet analysis},'' \emph{JOSS}, vol.~4, no.~36, p. 1237, 2019.

\bibitem{M4Singer}
L.~Zhang, R.~Li, S.~Wang, L.~Deng, J.~Liu, Y.~Ren, J.~He, R.~Huang, J.~Zhu, X.~Chen, and Z.~Zhao, ``{M4Singer: {A} Multi-Style, Multi-Singer and Musical Score Provided Mandarin Singing Corpus},'' in \emph{NeurIPS}, 2022.

\bibitem{PJS}
J.~Koguchi, S.~Takamichi, and M.~Morise, ``{PJS: phoneme-balanced Japanese singing-voice corpus},'' in \emph{{APSIPA}}, 2020, pp. 487--491.

\bibitem{Opencpop}
Y.~Wang, X.~Wang, P.~Zhu, J.~Wu, H.~Li, H.~Xue, Y.~Zhang, L.~Xie, and M.~Bi, ``{Opencpop: {A} High-Quality Open Source Chinese Popular Song Corpus for Singing Voice Synthesis},'' in \emph{{INTERSPEECH}}, 2022.

\bibitem{OpenSinger}
R.~Huang, F.~Chen, Y.~Ren, J.~Liu, C.~Cui, and Z.~Zhao, ``{Multi-Singer: Fast Multi-Singer Singing Voice Vocoder With {A} Large-Scale Corpus},'' in \emph{{ACM MM}}, 2021.

\bibitem{csd}
S.~Choi, W.~Kim, S.~Park, S.~Yong, and J.~Nam, ``{Children’s song dataset for singing voice research},'' in \emph{ISMIR}, 2020.

\bibitem{LibriTTS}
H.~Zen, V.~Dang, R.~Clark, Y.~Zhang, R.~J. Weiss, Y.~Jia, Z.~Chen, and Y.~Wu, ``{LibriTTS: {A} Corpus Derived from LibriSpeech for Text-to-Speech},'' in \emph{{INTERSPEECH}}, 2019, pp. 1526--1530.

\bibitem{ljspeech}
K.~Ito and L.~Johnson, ``{The LJ Speech Dataset},'' \url{https://keithito.com/LJ-Speech-Dataset/}, 2017.

\bibitem{vctk}
J.~Yamagishi, C.~Veaux, K.~MacDonald \emph{et~al.}, ``{CSTR VCTK Corpus: English multi-speaker corpus for CSTR voice cloning toolkit (version 0.92)},'' \emph{CSTR}, 2019.

\bibitem{adamw}
I.~Loshchilov and F.~Hutter, ``{Decoupled Weight Decay Regularization},'' in \emph{ICLR}, 2019.

\bibitem{APNet}
Y.~Ai and Z.~Ling, ``{APNet: An All-Frame-Level Neural Vocoder Incorporating Direct Prediction of Amplitude and Phase Spectra},'' \emph{TASLP}, pp. 2145--2157, 2023.

\bibitem{amphion}
X.~Zhang, L.~Xue, Y.~Wang, Y.~Gu, X.~Chen, Z.~Fang, H.~Chen, L.~Zou, C.~Wang, J.~Han, K.~Chen, H.~Li, and Z.~Wu, ``{Amphion: An Open-Source Audio, Music and Speech Generation Toolkit},'' \emph{arXiv}, vol. abs/2312.09911, 2023.

\bibitem{SVCC}
W.-C. Huang, L.~P. Violeta, S.~Liu, J.~Shi, Y.~Yasuda, and T.~Toda, ``{The Singing Voice Conversion Challenge 2023},'' \emph{arXiv}, vol. abs/2306.14422, 2023.

\bibitem{weightnorm}
T.~Salimans and D.~P. Kingma, ``{Weight Normalization: {A} Simple Reparameterization to Accelerate Training of Deep Neural Networks},'' in \emph{NeurIPS}, 2016, p. 901.

\bibitem{PESQ}
A.~W. Rix, J.~G. Beerends, M.~P. Hollier, and A.~P. Hekstra, ``{Perceptual evaluation of speech quality (PESQ)-a new method for speech quality assessment of telephone networks and codecs},'' in \emph{{ICASSP}}, 2001, pp. 749--752.

\bibitem{cargan}
M.~Morrison, R.~Kumar, K.~Kumar, P.~Seetharaman, A.~C. Courville, and Y.~Bengio, ``{Chunked Autoregressive {GAN} for Conditional Waveform Synthesis},'' in \emph{ICLR}, 2022.

\end{thebibliography}

\end{document}